\documentclass[prb,twocolumn,showpacs,amsmath,amssymb, floatfix,superscriptaddress]{revtex4}
\usepackage{graphicx}
\usepackage{dcolumn}
\usepackage{bm}

\newcommand{\ket}[1]{\left|#1\right>}
\newcommand{\bra}[1]{\left< #1 \right|}

\begin{document}

\title{Effects of fermions on the superfluid-insulator phase diagram of the Bose-Hubbard model}
\author{Sumanta Tewari}
\affiliation{Department of Physics and Astronomy, Clemson University, Clemson, SC 29634}
\author{Roman M. Lutchyn}
\affiliation{Condensed Matter Theory Center and Joint Quantum Institute, Department of Physics, University of
Maryland, College Park, MD 20742}

\author{S. Das~Sarma}
\affiliation{Condensed Matter Theory Center and Joint Quantum Institute, Department of Physics, University of
Maryland, College Park, MD 20742}
\date{\today}

\begin{abstract}
Building on the work of Fisher \textit{et al.} (Phys. Rev. B
  {\textbf{40}}, 546 (1989)), we develop the perturbation theory for the Bose-Hubbard model
  and apply it to calculate the effects of a degenerate gas of spin-polarized fermions interacting by contact
  interactions with the constituent bosons.
  For the single-band Bose-Hubbard model,
  we find that the net effect of the screening of the boson on-site interaction by the fermions is to suppress the Mott-insulating lobes in the Bose-Hubbard phase diagram.
  For the more general multi-band model, we find that, in addition to the fermion screening effects, the virtual excitations of the bosons to the higher Bloch bands, coupled with the contact interactions with the
  fermions, result in an effective increase (decrease) of the
boson on-site repulsion (hopping parameter). If the higher-band renormalization of the boson parameters is dominant over the fermion screening of the interaction, the Mott insulating lobes in the Bose-Hubbard phase diagram are enhanced for either sign of the Bose-Fermi interactions, consistent with the recent experiments.
\end{abstract}


\maketitle
\section{Introduction} The superfluid to insulator quantum phase transition in a degenerate gas of bosons moving in a periodic 
potential is described by the Bose-Hubbard model introduced by Fisher \textit{et al.}~[\onlinecite{Fisher_PRB89}] almost two decades ago. In its simplest
 form, the model includes the hopping term, $t$, which describes the nearest neighbor tunneling amplitude of the constituent bosons on the lattice, and the on-site repulsion term, $U$, which approximates the short-range part of the Coulomb interaction
 if the bosons are charged (contact interaction if the bosons are neutral). In addition, the model includes the boson chemical potential, $\mu$, which couples to the on-site charge density. The model can be directly implemented using spinless bosonic atoms moving in an artificially created periodic optical lattice. \cite{Jaksch, Bloch}

It is remarkable that a theory developed based on the above simple premises can describe a true quantum phase transition with enough predictive power that can be tested experimentally. \cite{Fisher_PRB89, Sachdev_book} For large repulsive
interaction $U$, boson charge fluctuations are suppressed and the
system is in an insulating state. On the other hand, when the
on-site repulsion is reduced, or, more appropriately, for large
$\frac{t}{U}$, the system is in a superfluid state due to
the Bose condensation of the mobile bosons. At some intervening value of $\frac{t}{U}$, then,
there should be a quantum phase transition separating the two
phases.~\cite{Fisher_PRB89, Sachdev_book} Recent
experiments~\cite{Greiner, Esslinger_PRL04, Spielman_PRL07,  Spielman_PRL08} using ultra cold bosonic atoms confined to an
optical lattice, which mimics the model for periodic external potential in a custom
setting, provided a first real demonstration of the superfluid to insulator transition in an experimental system.
The advantage of the atomic system lies in the ability to tune the parameters $t, U$ and $\mu$ at
will in a pristine, disorder-free environment.

Even for periodic, disorder-free, external potentials, in a real solid state
system additional fermions are always present and are invariably coupled to the constituent bosons.
For example, in a granular superconductor,~\cite{Beloborodov_review} where the Cooper pairs can be modeled
as the bosons, there can be thermally generated quasiparticles.~\cite{Lutchyn05} The question of additional fermions is also important in the context of the He$_3$-He$_4$ mixtures,~\cite{Bardeen_PRL66, Hemixtures_review} and the quark matter, where two (color) quarks form a Cooper pair which interacts with the remaining unpaired  quarks.~\cite{Shovkovy} Remarkably, such an additional degenerate gas of fermions can also be artificially introduced and coupled to the bosons in the ultra-cold atomic system.~\cite{Ferlaino} This raises an important theoretical question as to what happens to the phases of the original
Bose-Hubbard model \cite{Fisher_PRB89} when these additional fermions are present. The theoretical answer to this question can be experimentally tested in the so-called Bose-Fermi mixtures already realized in the optical lattice systems.~\cite{gunter_PRL06, ospelkaus_PRL06, Bloch08}

The effects of a degenerate gas of fermions on the Bose-Hubbard model have recently been investigated by
various theoretical methods.~\cite{Albus_PRA03, Roth_PRA04, Lewenstein_PRL04, Buchler_PRB04,Pollet_PRL06,
pollet_PRA08, Refael_PRB08, Mathey_PRL04, Mathey_PRB07, KSengupta_PRA07, PSengupta_PRA07, Mering_07, Scalettar, Tit, Lutchyn_PRB08, Lutchyn_PRA09, Adhikari, Frey, Powell} At first glance, the interaction between the bosons and the fermions seems to give rise to an effectively reduced repulsive interaction among the bosons compared to the bare (without the fermions) model. As a result, it may be expected that the superfluid phase coherence would increase in the Bose-Fermi mixture compared to the purely bosonic case. Recent theoretical and numerical works in Refs.~[\onlinecite{Buchler_PRB04,
pollet_PRA08}] seem to agree with this preliminary assertion. Quite surprisingly, however, the opposite effect -- the reduction of the superfluid coherence -- was observed in the experiments via the measurement of the visibility of the quasimomentum distribution.~\cite{gunter_PRL06, ospelkaus_PRL06, Bloch08}
The numerical work of Ref.~[\onlinecite{pollet_PRA08}] argues that the decrease of the bosonic phase coherence may actually be due to finite temperature effects.
\cite{cramer_PRL08}

The question of the intrinsic effect of the fermions on the Bose-Hubbard model is, however, far from theoretically settled. The assertion of the reduction of the repulsive interaction among the bosons is based on an hypothesis of static screening due to the fermions. \cite{Buchler_PRB04,pollet_PRA08} It has been recently argued in Ref.~[\onlinecite{Refael_PRB08}] that the dynamic part of the fermionic screening is equally important for the superfluid phase coherence of the bosons. By developing a Weiss-type, local, self-consistent mean field theory for this screening interaction, Ref.~[\onlinecite{Refael_PRB08}] claims that an effect akin to the fermionic orthogonality catastrophe, \cite{Anderson} arising from the fermionic dynamic screening fluctuations, can suppress superfluidity. This way, the net intrinsic effect of the fermions may be in the same direction as in the experiments after all.

In this work, we develop a rigorous perturbation theory to calculate the effects of an additional interaction potential (which, in the present context, is fermion-mediated) among the constituent bosons in the Bose-Hubbard model. To simplify the calculations, we take the system to be spatially homogenous, that is, we neglect the effects of the external confining potential in the optical set up. We also assume that the superfluid and the Mott insulating states are the only two possible states, and neglect the possibility of other exotic states.~\cite{Buchler_PRB04, KSengupta_PRA07, Tit, Adhikari, Frey,Lewenstein_PRL04, Powell,Mering_07} With these simplifying assumptions, we take the largest interaction to be the Hubbard $U$, and treat the additional fermion mediated interaction in perturbation theory. Using the Hubbard-Stratanovich transformation, we first rewrite the partition function of the model in terms of a space- and time-dependent complex scalar field theory, \cite{Fisher_PRB89, Sachdev_book} whose coupling constants are given by the correlation functions of the original Hamiltonian modified by the fermion-mediated interaction. In mean field theory, the coupling constant of the quadratic term of the field theory, which is given by the boson on-site Green's function, provides the phase boundary between the Mott-insulating and the superfluid phases. \cite{Sachdev_book} In the presence of the fermions, the boson Green's function must include the effects of the additional fermion-mediated interaction. The calculation of the phase boundary is thus reduced to the perturbative evaluation of the boson Green's function in the presence of the additional space-time-dependent interaction.

The above method still leaves us with a non-trivial problem, because the bare Bose-Hubbard model, on which we build the perturbation
expansion of the Green's function in powers of the additional interaction, is not Gaussian (quadratic in the boson operators). As a result, the standard machinery of bosonic perturbation expansion, \cite{Fetter_book} e.g., Wick's theorem and the linked cluster theorem which enable one to calculate the higher order corrections to the Green's function in terms of the integrals over products of the bare Green's function, do not apply. Thus, one has to calculate the higher-order correlation functions non-perturbatively with respect to the Hubbard $U$. Fortunately, we need these correlation functions computed only
with respect to the \textit{on-site} part of the Bose-Hubbard model, which conserves the number of particles on every site. Because of this local number conservation, we are able to calculate these correlation functions exactly in the particle number basis $\{n_i\}$. We also confirm that the apparent divergences, arising out of the summations over all the lattice sites and integrals over the imaginary time in the calculations of the correlation functions, are exactly canceled, and in the final result, one obtains non-zero perturbative corrections to the Green's function. Using this perturbatively corrected Green's function, we can calculate the effects of the fermions on the superfluid-insulator phase diagram.

Using the methods outlined above, we find that, for the single-band Bose-Hubbard model (sections \ref{section:1}, \ref{section:2}, \ref{section:3}, \ref{section:4}), the fermions
intrinsically shrink the area occupied by the Mott
insulating lobes (Fig.~\ref{fig:phase_boundary}). The overall effect is qualitatively in the same direction as in the
effects of Ohmic dissipation in enhancing the superconducting phase
coherence in Josephson junction array,~\cite{Chakravarty_PRB88, Tewari_PRB06} or
in granular superconductors.~\cite{Beloborodov_review} This result is contrary to the orthogonality
catastrophe argument of Ref.~[\onlinecite{Refael_PRB08}], while it agrees with the numerical results of Ref.~[\onlinecite{pollet_PRA08}].
Experiments, however, have quite convincingly shown that the fermions expand the area occupied by the Mott insulating lobes.
The earlier experiments \cite{gunter_PRL06,ospelkaus_PRL06} observed the loss of superfluid coherence for fixed attractive boson-fermion interactions, $U_{BF}$, which were larger in magnitude than the boson on-site repulsion itself. Recently, this finding has also been confirmed for both attractive and repulsive interspecies interactions in a range of values for $|U_{BF}|$ both smaller and larger than $U$. \cite{Bloch08} From our rigorous perturbation theory, therefore, we conclude that the single-band Bose-Hubbard model is inadequate to explain the experimentally observed
loss of superfluidity of the bosons by adding a degenerate gas of fermions.

   Next, we treat the more general multi-band Bose-Hubbard model in the presence of the fermions (section \ref{section:5}) in the same analytical framework developed for
    the single-band model. Here, we first find that there is an additional effect on the bosonic system, due to the fermion-boson contact interactions, which is mediated by virtual
   transitions of the bosons to the higher Bloch bands of the multi-band model. This effect leads to an effective \emph{increase} of the
   boson on-site repulsion, $U$, and a \emph{decrease} of the hopping parameter, $t$, for either sign of the fermion-boson interactions. There is some numerical evidence of this effect (termed self-trapping) in Ref.~[\onlinecite{luhmann-2007}], for the case of attractive interspecies interactions only.
    We treat the two disparate effects -- fermionic screening and the effects of the higher bands -- within the same analytical framework. This theory
  provides a consistent explanation of the loss of bosonic superfluid coherence by introducing fermions, irrespective of the sign of the interspecies interactions,
  as seen in the recent cold atom experiments. As a bonus, the perturbation theory we develop for the Bose-Hubbard model can be applied to
calculate the effects of any additional interaction (not necessarily fermion-mediated) on the Bose-Hubbard phase diagram.
For example, the effects of Ohmic dissipation, \cite{Dalidovich} or the effects of a second dilute gas of bosons on the superfluid-insulator phase diagram can also be evaluated by the methods described here. Some of our results were earlier presented in shorter forms in Refs.~[\onlinecite{Lutchyn_PRB08, Lutchyn_PRA09}].  We provide all the technical details of the theory in completeness in the current article, and mention that the results involving the renormalization of the bosonic hopping term due to the fermions in the presence of the multiband processes (section VI C), a mechanism of considerable quantitative importance in determining the final quantum phase diagram, was not considered earlier by us and is thus a completely new result.

\section{Single-band Bose-Hubbard model}
\label{section:1}
The Hamiltonian for the single-band Bose-Hubbard model is given by,
\begin{align}
H_B&=H_{\rm{os}}+H_t,\label{eq:Hb}\\
H_{\rm{os}}\!&=\!\sum_{i}\left(\frac{U}{2}
\hat{n}_{Bi}(\hat{n}_{Bi}-1)-\mu_{B}
\hat{n}_{Bi}\right),\label{eq:Hos}\\
H_t&=-t_B\sum_{<ij>}\left(b^{\dag}_i b_j
\!+\!H.c.\right).\label{eq:Ht}
\end{align}
Here $b^{\dag}_i, b_i$ are the spinless boson
creation and annihilation operators on the site $i$ and $\hat{n}_{Bi}=b_i^\dag b_i$ is the
boson density operator. $U>0$ and $\mu_{B}$ in the on-site part of the Hamiltonian, $H_{\rm{os}}$, denote the on-site boson-boson interaction and the bare chemical potential, respectively. The part of the Hamiltonian, $H_t$, that depends on the nearest neighbor pairs $\langle i,j \rangle$, involves the
nearest neighbor hopping matrix element, $t_B$, for the bosons.
The partition function for the model can be represented in terms of an imaginary-time path integral,
\begin{align}
Z_B=\int \mathfrak{D} b^{*} \mathfrak{D}b \exp\left[-S_B( b^{*}, b)\right],
\end{align}
where
\begin{align}
S_B( b^{*}, b)=\sum_i\int_{0}^{\beta} d\tau b^{*}_i \partial_{\tau} b_i+\int d\tau H_B,
\end{align}
and $\beta$ denotes the inverse temperature.

By decoupling the boson hopping term using the Hubbard-Stratanovich
transformation with a complex scalar field $\psi_i(\tau)$, the partition function becomes,
\begin{eqnarray}
Z_B&=&\int \prod_{i} \mathfrak{D} b_i^{*} \mathfrak{D} b_i
\mathfrak{D} \psi_i \!\mathfrak{D} \psi^*_i e^{-\int_{0}^{\beta} \! d
\tau \!\sum_{i,j}\!\psi^*_i(\tau)w^{-1}_{ij}\psi_j(\tau)\!}\nonumber\\ &\times& \exp\left(-S_{\rm
os}[b^{*}_i,b_i]+S_{c}[b^{*}_i,b_i,\psi^*_i,\psi_i]\right).
\end{eqnarray}
Here, the symmetric matrix $w_{ij}$ has non-zero elements, $t_B$, only for the nearest neighbors, $\langle i,j \rangle$.
The on-site part of the action, $S_{\rm{os}}$, is given by,
\begin{align}
\label{eq:Sos}
S_{\rm{os}}[b^{*}_i,b_i]&=\sum_i\int_{0}^{\beta} d\tau b^{*}_i \partial_{\tau} b_i\\
&+\int_{0}^{\beta} d\tau
\sum_{i}\left(\frac{U}{2} \hat{n}_{Bi}(\hat{n}_{Bi}-1)-\mu_B
\hat{n}_{Bi}\right).\nonumber
\end{align}
The coupled part of the action, $S_c$, is given by,
\begin{equation}
S_c[b^{*}_i,b_i,\psi^*_i,\psi_i]\!=\!\!\int_{0}^{\beta} \! d \tau
\!\sum_i\!\left(\psi^*_i(\tau)b_i(\tau)\!+\!\psi_i(\tau)b^{*}_i(\tau)\!\right)\!.
\end{equation}
  To write the effective theory in terms of the scalar field $\psi$ only, we perform the cumulant expansion,
\begin{align}
Z_B=\int \prod_{i}\mathfrak{D} \psi_i \mathfrak{D} \psi^*_i
&\exp\left(-\sum_i \frac{F_i(n_0,U,\mu)}{T}\right. \\
&\left. -\int_0^{\beta} d\tau d
\bm r \mathfrak{L}[\psi(\bm r,\tau),\psi^*(\bm r,\tau)]\right)\nonumber
\end{align}
where the Lagrangian
\begin{align} \label{eq:Phi4}
&{\cal{L}}[\psi(\bm r, \tau),\psi^*(\bm r, \tau)]=\\
&=\left(c_{1} \psi^* \frac{\partial\psi}{\partial
\tau}+c_{2} \left|\frac{\partial\psi}{\partial
\tau}\right|^2 +c \left|\bm \nabla{\psi}\right|^2 + r \left|\psi\right|^2 + u
\left|\psi\right|^4 \right)\nonumber
\end{align}
Here we used the continuum limit for $\psi_i(\tau)$, \emph{i.e.},
$\psi_i(\tau)\equiv\psi(\bm r,\tau)$. In this limit, the
coefficient $r$ is given by,
\begin{align}
r\propto\frac{1}{zt_b}-\int_{-\beta}^{\beta} d\tau \langle T_{\tau}
b_i(\tau) b^{\dagger}_i(0)\rangle
\label{eq:R}
\end{align}
where the brackets denote average with respect to the on-site part of the Hamiltonian $H_{\rm os}$.
Such imaginary-time-ordered averages can be conveniently calculated using the path integral
formalism with the on-site part of the action $S_{\rm{os}}$.
 In mean field theory, $r=0$ gives the
phase boundary between the insulator and the superfluid states.
Thus, the problem of calculating the phase diagram of the model is reduced to the calculation of the one-particle, on-site, boson Green's
function at zero Matsubara frequency $G_i(\omega_n=0)$, where $G_i(\tau \!-\! \tau^{\prime})=-\langle T_{\tau} b_i(\tau)b^{\dagger}_i(\tau^{\prime})\rangle$. The phase boundary is then determined by $r=0$:
\begin{align}
\frac{1}{zt_b}\!+\! G_i(0)\!=\!0,
\label{Eq:Phase-Boundary}
\end{align}
where $z$ is the coordination number of the lattice.
The on-site Green's function can be easily calculated using the boson number basis $\{n_i\}$,
\begin{align} \label{eq:green-time}
G_i(\tau\!-\!\tau')\!=\!-&\left[\Theta(\tau\!-\!\tau')\Theta(\delta E_p)(n_0\!+\!1)e^{-(\tau\!-\tau') \delta E_p} \right. \nonumber\\
\!&\!+\! \left.  \Theta(\tau'\!-\!\tau) \Theta(\delta E_h) n_0
e^{(\tau\!-\tau')\delta E_h}\right].
\end{align}
Here $n_0$ is the mean density of the bosons in the ground state at zero temperature, $\delta E_p=U n_0-\mu$ and $\delta E_h=\mu - U (n_0-1)$ are the particle and hole excitation energies, respectively.
In the frequency domain, this function becomes, \cite{Sachdev_book}
\begin{equation}
G_i(i\omega_n)\!=\left[\frac{(n_0+1)\Theta[-\mu+U
n_0]}{i\omega_n\!+\!\mu-Un_0}\!-\!\frac{n_0
\Theta[\mu\!-\!U(n_0\!-\!1)]}{i\omega_n\!+\!\mu-U(n_0\!-\!1)}\right]\!.
\label{Eq:Green-Frequency}
\end{equation}
Finally, using Eqs.~(\ref{Eq:Phase-Boundary}) and (\ref{Eq:Green-Frequency}) one arrives at the generic Bose-Hubbard phase diagram on the $(\mu, t_B)$ plane as shown by the solid curve in Fig.~\ref{fig:phase_boundary}.

\section{Single-band Bose-Hubbard model with fermions}
\label{section:2}
\subsection{Partition function}
We consider a mixture of bosonic and spin-polarized fermionic atoms
in an optical lattice. The full Hamiltonian of the Bose-Fermi system is given by
$H=H_B+H_F+H_{BF}$, with $H_F$ representing the
fermionic part of the Hamiltonian and $H_{BF}$ describing the
inter-species interaction:
\begin{align}
&H_F\!=\!-t_F\sum_{<ij>}\left(c^{\dag}_i c_j
\!+\!H.c.\right)-\mu_{F}\sum_i c^{\dag}_i c_i,\label{eq:Hf}\\
&H_{BF}\!=\!U_{FB}\sum_i \hat{n}_{Bi} (c^{\dag}_i c_i-n^0_{Fi}).
\label{eq:Hfb}
\end{align}
Here $c^\dag_i, c_i$ are the fermion
creation and annihilation operators on site $i$, $t_F$ corresponds to the nearest neighbor
hopping matrix element for the fermions, $\mu_F$ is the fermion chemical potential, $U_{FB}$ describes the on-site
boson-fermion interaction,  and $n^0_{Fi}$ is the
average density of the fermions.  In Eq.~(\ref{eq:Hfb}), the quantity $n^0_{Fi}$ has been subtracted
from the fermionic density, $c^{\dag}_ic_i$, to highlight the
lowest order (in $U_{FB}$) effect of the fermions on the
constituent bosons, which is a trivial shift of the boson chemical
potential: $\mu_B \rightarrow\mu_B-U_{FB}n^0_{Fi}$. Henceforth, this shift in the chemical potential is
implicitly assumed in $H_B$.

The partition function of the Bose-Fermi system is given as,
\begin{align}
Z=\int \mathfrak{D} b^{*} \mathfrak{D} b \mathfrak{D} c^{\dag}
\mathfrak{D} c \exp\left[-S( b^{*}, b, c^{\dag},c)\right].
\end{align}
Here, the action $S( b^{*}, b, c^{\dag},c)$ is given by
\begin{align}
S( b^{*}, b, c^{\dag},c)=\sum_i\int_{0}^{\beta} d\tau (b^{*}_i \partial_{\tau} b_i+c^{\dag}_i
\partial_{\tau} c_i)+\int_{0}^{\beta} d\tau H.
\end{align}
To write an effective field theory analogous to that in Eq.~(\ref{eq:Phi4}) we need to successively integrate out
the fermions and the bosons as discussed below.

\subsection{Integrating out the fermions}
The first non-trivial effects due to the fermions
appear in the second order in $U_{FB}$. By integrating out the fermions
(note
that the fermions appear only in quadratic order in $H$),
the imaginary-time partition function becomes,
\begin{align}
Z&=\int \mathfrak{D} b_i^{*} \mathfrak{D} b_i \exp\left(-S_{\rm eff}[ b_i^{*}, b_i]\right) \label{eq:part}\\
S_{\rm eff}[b_i^{*},b_i]&=\int_{0}^{\beta} d\tau \sum_i b^{*}_i \partial_{\tau} b_i+\int_{0}^{\beta}d\tau \left(H_{\rm{os}}+H_{t} \right)\label{eq:eff-action}\nonumber\\
&-\sum_{ij} \int_0^{\beta} d \tau_1 \int_0^{\beta} d \tau_2
n_{Bi}(\tau_1)M_{ij}(\tau_1\!-\!\tau_2) n_{Bj}(\tau_2).\nonumber\\
\end{align}
In the second order in $U_{FB}$, the integral over the fermion degrees of freedom gives rise to an effective
non-local density-density interaction for the bosons with the function $M_{ij}(\tau_1-\tau_2)$ defined as,
\begin{equation}
M_{ij}(\tau_1-\tau_2)=\frac{U_{FB}^2}{2}\left\langle \Delta n_{Fi}(\tau_1) \Delta n_{Fj}(\tau_2) \right\rangle.\label{eq:M(ij)}
\end{equation}
In the frequency and momentum domain, the effective interaction $M_{\bm q}(\Omega_n)$ is proportional to the fermion polarization function. The exact form of the interaction $M_{\bm q}(\Omega_n)$ depends on the dimensionality of the system. The effective fermion-mediated boson-boson interaction kernel in 3D is given by,
\begin{widetext}
\begin{align}\label{eq:polarization3D}
M_{\bm q}(\Omega_n)\!=\!\frac{U_{FB}^2}{2\Delta}
\left( \frac{1}{2}+\frac{1}{8k} \left[1-\left( k-\frac{i\nu_n}{k} \right)^2 \right]
\ln\! \left[ \frac{k-i\frac{\nu_n}{k}+1}{k-i\frac{\nu_n}{k}-1} \right]+\frac{1}{8k} \left[1\!-\!\left( k+\frac{i\nu_n}{k} \right)^2 \right]
\ln\! \left[ \frac{k+i\frac{\nu_n}{k}+1}{k+i\frac{\nu_n}{k}-1} \right]  \right),
\end{align}
\end{widetext}
Here, $\nu_n=\Omega_n/4E_F$ and $k=q/2k_F$, with $E_F$ and $k_F$ being the Fermi energy and the Fermi momentum, respectively. $\Delta$ is the fermion mean level-spacing, $\Delta=1/\nu_FV$, with $\nu_F$ the density of states at the Fermi level and $V$ the volume of the unit cell, $V=a^3$.

\subsection{Integrating out the bosons}
 Using the Hubbard-Stratanovich transformation, we first decouple the boson hopping term, $H_t$, in Eq.~(\ref{eq:eff-action}). We then integrate out the bosonic fields via cumulant expansion to find,
\begin{equation}
Z=Z_0 \int \mathfrak{D} \psi_i  \mathfrak{D} \psi^*_i \exp(-S[\psi_i,\psi_i^*]),
\end{equation}
where the action $S[\psi_i, \psi_i^*]$ is given by,
\begin{align}\label{Spsi}
S[\psi_i, \psi_i^*]&=\int_0^{\beta} d \tau \sum_{i,j} \psi^*_i(\tau)
w^{-1}_{ij} \psi_j(\tau)\\\nonumber &- \ln \! \left  \langle \exp
\left[ \int_0^{\beta} d \tau \sum_i b_i(\tau)
\psi_i^*(\tau)+ H. c. \right] \! \right \rangle^{\prime} \!.
\end{align}
The
expectation value $\langle ... \rangle^{\prime}$ in Eq.~(\ref{Spsi}) is taken with respect to the action $S_{\rm eff}[b_i^{*},b_i]$ with the boson hopping parameter $t_B=0$, i.e., with respect to the action
\begin{equation}
\label{eq:S'}
S^{\prime}=S_{\rm{os}}-\sum_{ij} \int_0^{\beta} d \tau_1 \int_0^{\beta} d \tau_2
n_{Bi}(\tau_1)M_{ij}(\tau_1\!-\!\tau_2) n_{Bj}(\tau_2).
  \end{equation}
  By expanding $S[\psi, \psi^*]$ up to the fourth power of the field $\psi$, and taking the continuum limit, we arrive at the action of an effective complex $\phi^4$ field theory,
\begin{equation}
\!\!{\cal{S}}[\psi,\!\psi^*]\!\!=\!\!\!\int \!\! d\bm x  \!\! \left(\!\!c'_{1} \psi^* \frac{\partial\psi}{\partial
\tau}+\!c'_{2}\! \left|\frac{\partial\psi}{\partial
\tau}\right|^2\!\!+\! c'\! \left|\bm \nabla{\psi}\right|^2\!+\!r'\! \left|\psi\right|^2\!+\!u'\!
\left|\psi\right|^4\!\right)\!\! \label{eq:Phi'4}
\end{equation}
with $\bm x=\{\bm r,\tau \}$. The coupling constants $c'_{1},c'_2, c', r', u'$
are given by the correlation functions of the bosonic fields with respect to the action $S'$.
As before, the mean-field phase boundary between the
superfluid and insulating phases can be obtained by
setting the coefficient $r'$ to zero:
\begin{align}\label{eq:r}
r' \propto \frac{1}{z t_B}+\int_{-\beta}^{\beta} d\tau {{G}}^{\prime}_i(\tau)=0,
\end{align}
where $ {{G}}^{\prime}_i(\tau)=-\langle T_{\tau} b_i(\tau)b_i^{\dagger}(0)\rangle^{\prime}$ is the single-site boson Green's function, which, in the presence of the fermions, must now include the effects of the additional fermion-mediated density-density interaction.
Thus, the problem is now reduced to the calculation of
the on-site full boson Green's function by computing the corrections
to Eq.~(\ref{eq:green-time}) due to the fermion mediated interaction. As we show in the next section, this can be done perturbatively in $U_{FB}$.

\section{Boson Green's function in the presence of fermions}
\label{section:3}
The calculation of the perturbative corrections to the bare boson Green's
function, Eq.~(\ref{eq:green-time}), is non-trivial because the bare
on-site Hamiltonian, $H_{\rm{os}}$, is not quadratic in the boson operators. Therefore, one
cannot use the standard diagrammatic technique, \cite{Fetter_book}
because the Wick's theorem does not hold. To make progress, we write the average required for the corrected
Green's function as,
\begin{eqnarray}
\label{eq:expansion}
\langle T_{\tau} b_i(\tau)b_i^{\dagger}(0)\rangle^{\prime}&=&
\frac{1}{Z^{\prime}}\int\mathfrak{D}b^{*}\mathfrak{D}b \exp(-S^{\prime})b_i(\tau)b_i^{*}(0)\\
&=&\frac{1}{Z^{\prime}}\int\mathfrak{D}b^{*}\mathfrak{D}b \exp(-S_{\rm{os}}+\lambda S_1)b_i(\tau)b_i^{*}(0),\nonumber
\end{eqnarray}
where $Z^{\prime}$ is the partition function corresponding to the action $S^{\prime}$ in Eq.~(\ref{eq:S'}), which we have rewritten here using the definition
\begin{equation}\lambda S_1=\sum_{ij} \int_0^{\beta} d \tau_1 \int_0^{\beta} d \tau_2
n_{Bi}(\tau_1)M_{ij}(\tau_1\!-\!\tau_2) n_{Bj}(\tau_2),
\end{equation}
where $\lambda$ is used as a book-keeping parameter. Note that, from this definition, the linear order in $\lambda$ corresponds to the quadratic order in the boson-fermion coupling constant, $U_{FB}$, via Eq.~(\ref{eq:M(ij)}). Evaluating the
perturbative corrections to the boson Green's function up to the quadratic order in $U_{FB}$, therefore, requires us to expand the second line of Eq.~(\ref{eq:expansion}) up to the linear order in $\lambda$. Expanding the numerator and the denominator of Eq.~(\ref{eq:expansion}) and collecting terms which are linear order in $\lambda$, we get,
\begin{align}\label{eq:expansion2}
\!\langle T_{\tau}b_i(\tau)b_i^{\dagger}(0)\rangle^{\prime}\!&=\!\langle T_{\tau} b_i(\tau)b_i^{\dagger}(0)\rangle \!\\&+\! \lambda \langle T_{\tau}b_i(\tau)b_i^{\dagger}(0)S_1\rangle \!-\! \lambda \langle T_{\tau} b_i(\tau)b_i^{\dagger}(0)\rangle
\langle S_1 \rangle. \nonumber
\end{align}
Finally, putting $\lambda =1$ in Eq.~(\ref{eq:expansion2}), we derive,
\begin{align}\label{eq:Green's}
\! \langle T_{\tau}
b_i(\tau)b^{\dagger}_i(0) \rangle \! ^{\prime}&\!=\!\!\langle T_{\tau}
b_i(\tau)b^{\dagger}_i(0) \rangle\\
\!&\!+\!\!\sum_{jl}\!\int_{-\beta}^{\beta} \!\! d \tau_1\!\! \int_{-\beta}^{\beta} d \tau_2
M_{jl}(\tau_1\!-\!\tau_2)K_{ijl}(\tau,\tau_1,\tau_2)\nonumber.
\end{align}
The higher-order correlation function $K_{ijl}(\tau,\tau_1,\tau_2)$ can be conveniently calculated using second quantization representation (see Appendix~\ref{app:correlation}), in which it is defined as
\begin{align}\label{eq:K-def}
K_{ijl}(\tau,\tau_1,\tau_2)=&\langle T_{\tau}
b_i(\tau)b^{\dag}_i(0) n_{j}(\tau_1) n_{l}(\tau_2) \rangle\nonumber\\
-&\langle T_{\tau}
b_i(\tau)b^{\dag}_i(0) \rangle \langle  T_{\tau} n_{j}(\tau_1) n_{l}(\tau_2) \rangle.
\end{align}
Thus, we have reduced the problem of calculating the perturbative corrections to the boson Green's function due to the fermion-mediated, non-local interaction  to calculating a higher order boson correlation function
with respect to $H_{\rm{os}}$ given in Eq.~(\ref{eq:Hos}).

\subsection{Higher order boson correlation functions}
Since the on-site part of the Bose Hubbard Hamiltonian conserves
the number of bosons, the correlation functions above can be
calculated exactly using the particle-number eigenstates. From Eq.~(\ref{eq:K-def}), we see that there
are two free spatial indices $j,l$  and two free imaginary time indices $\tau_1, \tau_2$ in the definition
of $K_{ijl} (\tau, \tau_1, \tau_2)$. It is clear from Eq.~(\ref{eq:Green's}) that the correction to the bare
Green's function involves sums over the free spatial indices and integrals over the free imaginary times, both of which, in principle, can give diverging contributions (note that $\beta \rightarrow \infty$ as $T\rightarrow 0$). However, as we show in Appendix ~\ref{app:correlation}, the second term on the right hand side of Eq.~(\ref{eq:K-def}) subtracts these diverging terms exactly by constraining the spatial sums to $i$ and limiting the $\tau_1, \tau_2$ integrals to the imaginary time intervals between $0$ and $\tau$. In this way, we are able to get rid of the divergences in the perturbation theory, even though the usual linked cluster theorem \cite{Fetter_book} does not apply to the interacting boson Hubbard model.
After subtracting the terms which would have produced divergent contributions in Eq.~(\ref{eq:Green's}), the correlation function $K_{ijl}(\tau, \tau_1, \tau_2)$ acquires the following form at $T=0$:
\begin{widetext}
\begin{align}\label{eq:K-calc}
&K_{ijl}(\tau, \tau_1, \tau_2)=e^{-\delta E_p \tau}\Theta(\delta E_p) \Theta(\tau) \left \{ \Theta (\tau\!-\!\tau_1)\Theta (\tau\!-\!\tau_2)\Theta (\tau_2)\Theta (\tau_1)\left[(n_0\!+\!1)n_0(\delta_{ij}\!+\!\delta_{il})\!+\!(n_0\!+\!1)\delta_{ij}\delta_{il}\right] \right. \\
&\!+\Theta (\tau_2\!-\!\tau)\Theta (\tau\!-\!\tau_1)\Theta (\tau_2)\Theta (\tau_1)\delta_{ij}n_0(n_0\!+\!1)\!+\!\Theta (\tau_1\!-\!\tau)\Theta (\tau\!-\!\tau_2)\Theta (\tau_2)\Theta (\tau_1)\delta_{il}n_0(n_0\!+\!1)\nonumber\\
&\left.\!+\Theta (\tau\!-\!\tau_1)\Theta (\!-\!\tau_2)\Theta (\tau_1)\delta_{ij}n_0(n_0\!+\!1)\!+\!\Theta (\tau\!-\!\tau_2)\Theta (\tau_2)\Theta (\!-\!\tau_1)\delta_{il}n_0(n_0\!+\!1) \right \}\nonumber\\
&\!-\!e^{\delta E_h \tau} \Theta(\delta E_h) \Theta(-\tau) \left \{\Theta (\tau_1\!-\!\tau)\Theta (\tau_2)\Theta (\!-\!\tau_1) n_0^2\delta_{ij}+\Theta (\tau_2\!-\!\tau)\Theta (\!-\!\tau_2)\Theta (\tau_1)n_0^2\delta_{il}+\!\Theta (\tau_2\!-\!\tau)\Theta (\tau\!-\!\tau_1)\Theta (\!-\!\tau_2)\Theta (\!-\!\tau_1)n_0^2\delta_{il} \right. \nonumber\\
&\!+\left. \Theta (\tau_1\!-\!\tau)\Theta (\tau\!-\!\tau_2)\Theta (\!-\!\tau_2)\Theta (\!-\!\tau_1)n_0^2\delta_{ij}+\!\Theta (\tau_1\!-\!\tau)\Theta (\tau_2\!-\!\tau)\Theta (\!-\!\tau_2)\Theta (\!-\!\tau_1) [n_0^2(\delta_{ij}\!+\!\delta_{il})\!-\!\delta_{ij}\delta_{il}n_0]\right \}.\nonumber
\end{align}
\end{widetext}
Here $\delta E_p$ and $\delta E_h$ are the particle and the hole excitation energies: $ \delta E_p={U n_0 -\mu}$ and $\delta
E_h={\mu -U (n_0 -1)}$.
It is important to note that the correlation function $K_{ijl}(\tau, \tau_1, \tau_2)$ is irreducible and cannot be factored into the product of the bare Green's functions as would have been possible if Wick's theorem were applicable.

\subsection{Green's function in static approximation}
We now proceed to calculate the effects of the fermions on the boson Green's function
in the static approximation (static limit of the fermion polarization function $\Omega_n=0$ and $\bm q \rightarrow 0$) neglecting the retardation effects of the induced interaction among the bosons. The expression for $M_{\bm q}(\Omega_n)$ in Eq.~(\ref{eq:polarization3D}) becomes $M_{\bm q}(\Omega_n)\sim\frac{U_{FB}^2}{2\Delta}$.~\cite{Buchler_PRB04, Refael_PRB08} We then substitute
the corresponding expression for $M_{jl}$,
$M_{jl}(\tau_1-\tau_2)=\frac{U_{FB}^2}{2\Delta}\delta_{jl}\delta(\tau_1-\tau_2)$,
into the correction term to the bare Green's function, which is the second term in Eq.~(\ref{eq:Green's}).
Because of the factor $\delta(\tau_1 - \tau_2)$ in $M_{jl}(\tau_1 -\tau_2)$,
we note that the correlation function $K_{ijl}(\tau, \tau_1, \tau_2)$ in Eq.~(\ref{eq:K-calc}) reduces, in the static approximation, to
\begin{align}\label{eq:K-calc_static}
\! & K_{ijl}(\tau,\tau_1,\tau_2)
\!\rightarrow
\Theta(\tau)\Theta(\tau_1)\Theta(\tau_2)\Theta(\tau\!-\!\tau_1)\Theta(\tau\!-\!\tau_2)\nonumber\\
& \times
\left[(\delta_{ij}\!+\!\delta_{il})n_0(n_0\!+\!1)\!+\!\delta_{ij}\delta_{il}(n_0\!+\!1)\right]\exp \left(-\delta
E_p\tau\right)\nonumber\\
& +
\Theta(-\tau)\Theta(-\tau_1)\Theta(-\tau_2)\Theta(\tau_1-\tau)\Theta(\tau_2-\tau)\nonumber\\
& \times \left[-(\delta_{ij}+\delta_{il})n_0^2+\delta_{ij}\delta_{il}n_0\right]\exp\left(\delta
E_h\tau\right).
\end{align}
Substituting this form of $K_{ijl}$ into the correction term to the bare Green's function, using
the spatial Kronecker $\delta$-function from the kernel $M_{jl}$, and using the identity
$\sum_{j,l}(\delta_{ij}+\delta_{il})=2\sum_{j,l}\delta_{ij}$, we find the correction in the static approximation to be given by,
\begin{eqnarray}
\delta G_{i}^{{\prime \rm{s}}}(\tau)&=&
-\frac{U_{FB}^2}{2\Delta}\left(\theta(\tau)\int_{0}^{\tau}d\tau_1(n_0+1)(2n_0+1)e^{-\delta E_p\tau}\right. \nonumber\\
&+& \left. \theta(-\tau)\int_{\tau}^{0}d\tau_1 n_0 (1-2n_0)e^{\delta E_h\tau}\right).
\end{eqnarray}
Carrying out the $\tau_1$ integral and taking the Fourier transform to frequency space with respect to $\tau$,
we finally derive the following expression for the full Green's function in the static approximation
at zero frequency,
\begin{align}\label{eq:Green's-stat}
&{G}^{\prime\rm{s}}_i(0)= - \frac{1}{U} \left ({n_0+1 \over n_0- \mu/U}+{n_0 \over  \mu/U - (n_0-1)} \right) \\
&- {U_{FB}^2 \over 2 \Delta U^2} \left( {(n_0+1)(2n_0+1) \over (n_0- \mu/U)^2} -{n_0(2n_0-1) \over (\mu/U-(n_0-1))^2} \right)\nonumber.
\end{align}
It is clear from this expression that near the degeneracy points ($\frac{\mu}{U}$ is an integer and $t_B=0$), where the gap in the Mott insulator states to the single particle excitations, $\delta
  E_{p/h}$, is small  $\sim U_{FB}^2/\Delta$, the perturbation theory breaks down. This caveat would apply also to the calculation of the full Green's function given in the next subsection, and our results for the phase diagram would not be valid near the degeneracy points.

 In the static approximation, the same corrected Green's function as above could be obtained in an independent, more
obvious way. The method described above is useful, however, to calculate the corrections to the Green's function
when the perturbing term to the Hamiltonian is explicitly dependent on space and imaginary time, as in the case of the full fermion-mediated interaction. As we show below, the alternative, more transparent method to evaluate the Green's function in the static approximation produces the same answer as that found above, and this demonstrates the correctness of our perturbation formalism. In the alternative method, we could substitute the static, on-site form of $M_{ij}(\tau_1-\tau_2)$ directly into the action,
Eq.~(\ref{eq:eff-action}), and calculate the Green's function
exactly. It is easy to see that, in the static approximation, the mobile fermions simply renormalize $\mu$ and $U$ of the bare Bose Hubbard Hamiltonian $H_B$: $U\rightarrow U- U_{FB}^2/\Delta$ and $\mu\rightarrow \mu+U_{FB}^2/2\Delta$. The exact Green's function, thus, can simply be obtained by substituting these renormalized parameters in Eq.~(\ref{Eq:Green-Frequency}). After expanding the result to the second order in $U_{FB}$, the resulting expression exactly matches that in Eq.~(\ref{eq:Green's-stat}).

\subsection{Full boson Green's function}
The static screening approximation for $M_{\bm q}(\Omega_n)$ does not, however, take into account the important retardation effects \cite{Refael_PRB08} and the spatially non-local nature of the interaction kernel in Eq.~(\ref{eq:M(ij)}). In order to take into account these effects, we substitute the full expression for $M_{ij}(\tau_1-\tau_2)$ into Eq.~(\ref{eq:Green's}). Calling the second term on the right hand side of Eq.~(\ref{eq:Green's}) $\delta G_i^{\prime}(\tau)$, we find the correction to the bare Green's function to be given by,
\begin{eqnarray}
\delta G_i^{\prime}(\tau)&=&\left(\Theta(\tau)(n_0+1)e^{-\delta
E_p(\tau)}+\Theta(-\tau)n_0e^{\delta E_h(\tau)}\right)\nonumber\\&\times&
\int_{0}^{\tau} d \tau_1 \int_{0}^{\tau} d \tau_2
M_{ii}(\tau_1\!-\!\tau_2)
\end{eqnarray}

After performing the imaginary time integrals, we find that the required correction
to the Green's function at zero frequency (see Eq.~(\ref{eq:r})) is given by,
\begin{figure}[t]
\centering
\includegraphics[width=0.9\linewidth]{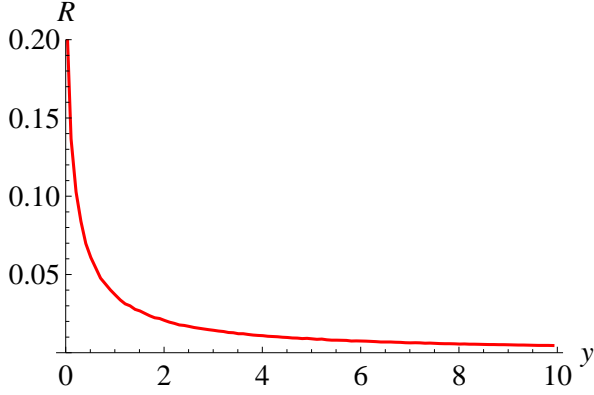}
  \caption{(color online) The dependence of the function $R^{}(y)$ on its argument.} \label{fig:Ry}
\end{figure}
\begin{eqnarray}
\int_{-\beta}^{\beta} d\tau
\delta G_i^{\prime}(\tau)&=&\frac 1 \beta \sum_{n=-\infty}^{n=\infty}
M_{ii}(\omega_n)\frac {4} {\omega_n^2} \int_{-\beta}^{\beta}d \tau
\sin^2\left[\frac{\omega_n
\tau}{2}\right]\nonumber\\ &\times&[\Theta(\tau)(n_0+1)e^{-\delta E_p
\tau}+\Theta(-\tau)n_0 e^{\delta E_h\tau}]\nonumber\\
&=&2\int_{-\infty}^{\infty}\frac{d\omega_n}{2\pi}\!M_{ii}(\omega_n)\nonumber\\ &\times&
\!\left[\frac{n_0\!+\!1}{\delta E_p}\frac{1}{\omega_n^2+\delta
E_p^2}\!+\!\frac{n_0}{\delta E_h}\frac{1}{\omega_n^2\!+\!\delta
E_h^2}\right]
\end{eqnarray}

Finally, using the form of $M_\mathbf{q}(\Omega_n)$ from Eq.~(\ref{eq:polarization3D}), we obtain the following expression for the full boson Green's function at zero frequency,
\begin{align}\label{eq:Green's_dyn}
{G}^{\prime}_i(0)=& - \frac{1}{U} \left( {n_0+1 \over n_0- \mu/U}+{n_0 \over  \mu/U - (n_0-1)} \right)  \\
&-{U_{FB}^2 \over  \Delta U^2} \left \{ {(n_0\!+\!1) \over (n_0\!-\!\mu/U)^2}R^{}\!\left(\frac{U}{4E_F}\left[n_0\!-\!\frac \mu U\right]\right) \right.\nonumber \\
&\!+\! \left. {n_0 \over (\mu/U\!-\!(n_0\!-\!1))^2} R^{}\!\left(\frac{U}{4E_F}\left[\frac \mu U \!-\!(n_0\!-\!1)\right]\! \right) \! \right \}\! \nonumber.
\end{align}
Here we introduced the dimensionless function $R^{}(y)$ given by,
\begin{widetext}
\begin{align}
R^{}(y)\!=\!\frac{1}{2\Lambda^{3}}\!\int_0^{\Lambda}\!k^2dk\!\int_0^{\infty}\!d\nu \!\left( \frac{1}{2}\!+\!\frac{1}{8k}\!\left[1\!-\!\left( k\!-\!\frac{i\nu}{k}\! \right)^2\! \right]\!
\ln\! \left[ \frac{k\!-\! i\frac{\nu}{k}\!+\!1}{k\!-\! i\frac{\nu}{k}\!-\!1} \right]\!+\!\frac{1}{8k} \left[1\!-\!\left( k\!+\!\frac{i\nu}{k} \right)^2\! \right]
\ln\! \left[ \frac{k\!+\!i\frac{\nu}{k}\!+\!1}{k\!+\! i\frac{\nu}{k}\!-\!1}\! \right]\!  \right)\frac{y}{\nu^2+y^2},
\end{align}
\end{widetext}
where the momentum integral is defined in the First Brillouin zone with $\Lambda=\frac{\pi}{2k_F a}$. The plot of the function $R^{}(y)$ is shown in Fig.~\ref{fig:Ry}.
As follows from Eq.~(\ref{eq:Green's_dyn}), the importance of the fermion renormalization effects is determined by the ratio of $\mu (\sim U)$ and $E_F$.
When the fermion density is small, \emph{i.e.}, $\mu/E_F\gg 1$, the corrections to the Green's function are suppressed since $R (y\gg 1)\rightarrow 0$. In the opposite limit, $\mu/ E_F\ll 1$, the function $R(y\ll 1)\sim 1$, and thus, for a given value of $U_{FB}$, the effects of the fermions on the bosons are more pronounced.

\section{Shift of the phase diagram within single-band model}
\label{section:4}
\begin{figure}{t}
\centering
\includegraphics[width=0.9\linewidth]{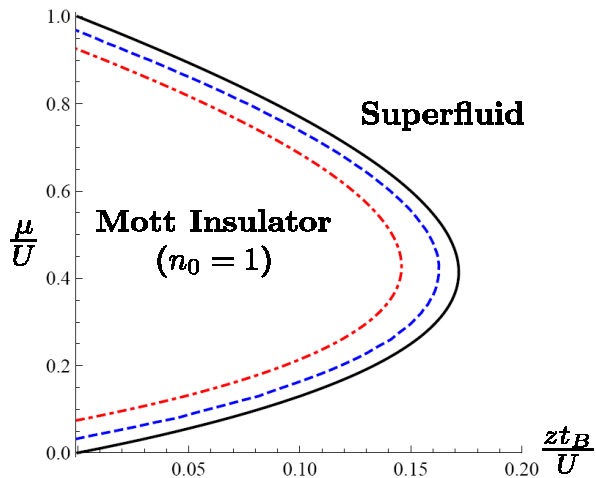}
  \caption{(color online) Phase diagram of the Bose-Hubbard model with and without the fermions in 3D for the boson density $n_0=1$. The phase diagram in 2D \cite{Lutchyn_PRB08} is qualitatively similar. Solid line describes the
insulator-superfluid phase boundary without the fermions. The dashed line corresponds to the same phase boundary when the effects of the
fermions are taken into account. For comparison, we also show the dash-dot line which denotes the phase boundary when the effects of the fermions are only treated in the static approximation, see text for details. The region near the degeneracy points (integer $\frac{\mu}{U}$), where the calculations of this paper do not apply, are implicitly excluded from this figure, see also discussion following Eq.(\ref{eq:Green's-stat}).
Here we used, for illustrative purposes, ${U \over 4E_F}\!=\!0.1$ and $\frac{U_{FB}^2}{\Delta
U}\!=\!0.15$.} \label{fig:phase_boundary}
\end{figure}

As we discussed in section \ref{section:2}, using Eqs.~(\ref{Eq:Phase-Boundary}, \ref{Eq:Green-Frequency}) which are applicable to the bare Bose-Hubbard model, one arrives at the bare Bose-Hubbard phase diagram on the $(\mu, t_B)$ plane as shown by the solid curve in Fig.~\ref{fig:phase_boundary}. Here we have plotted the figure for 3D, while the figure in two dimensions \cite{Lutchyn_PRB08} is qualitatively the same.  Even though, in this figure, we have shown the phase boundary between the Mott insulating and the superfluid phases
for only the mean ground state boson density $n_0=1$, the results for other integer boson densities are qualitatively
the same, \cite{Fisher_PRB89, Sachdev_book} and we have omitted them for simplicity. The Mott insulating states are characterized by $r> 0, \langle \psi \rangle = 0$ and survive in lobes extending from one degeneracy point to the next. The superfluid state, on the other hand, is characterized by a non-zero value of the order parameter: $r<0, \langle \psi\rangle \neq 0$. The transition between the Mott insulating and the superfluid states is a continuous quantum phase transition which can be described by the $\phi^{4}$ field theory given in Eq.~(\ref{eq:Phi4}).

In the static approximation for the screening, only the instantaneous part of the fermion polarization function is taken into account. The corresponding phase boundary is given by the dash-dot line in
Fig.~\ref{fig:phase_boundary}.
One can see that, in the static approximation, the fermions markedly shrink the area of the Mott-insulating lobes in the phase diagram.
 Finally, using Eq.~(\ref{eq:r}) and Eq.~(\ref{eq:Green's_dyn}), we calculate the true phase boundary, within the single-band Bose-Hubbard model, as shown by the dashed line in Fig.~\ref{fig:phase_boundary}. It is clear from this plot that the dynamic part of the fermion screening function indeed suppresses superfluidity, as argued in Ref.~[\onlinecite{Refael_PRB08}]. However, the net effect of the fermions, in a generic region away from the degeneracy points (where our calculations do not apply), is to still suppress the Mott-insulating lobes and enhance the area occupied by the superfluid state. The sign of this overall effect is qualitatively the same as in the
effects of Ohmic dissipation in enhancing the superconducting phase
coherence in Josephson junction array,~\cite{Chakravarty_PRB88, Tewari_PRB06} or
in granular superconductors.~\cite{Beloborodov_review} The single-band Bose-Hubbard model, therefore, proves to be inadequate in describing the loss
of superfluid coherence by adding fermions as seen in the experiments. In the next section, we will extend the analysis to the multi-band model in search of a consistent explanation of the experiments.

\section{Multi-band Bose-Hubbard model with fermions}
\label{section:5}

\subsection{Effective three-band model for Bose-Fermi mixtures}

As we have seen in the last section, the effects of fermions on the phase diagram of the single-band Bose-Hubbard model is in a direction opposite
to that seen in the experiments. It is then clear that the effects of the higher boson Bloch bands, which are so far neglected in the single-band model but can be significant in the experiments, should be taken into account.
    For the more general multi-band Bose-Hubbard model, there is an additional, higher-band, effect of the fermion contact interactions, which leads to an effective \emph{increase} of the
   boson on-site repulsion and a \emph{decrease} of the hopping parameter. As we show below, when this higher-band effect dominates over the fermion screening of the bosonic interactions,
  it provides an explanation for the loss of bosonic superfluid coherence by introducing fermions \emph{irrespective of the sign} of the interspecies interactions. \cite{gunter_PRL06, ospelkaus_PRL06, Bloch08}

To elucidate the effects of the higher boson Bloch bands, let's start with the following second quantized Hamiltonian:
\begin{widetext}
\begin{align}
H&=\int d^d \bm r \Phi^\dag (\bm r)\left[\frac{\hat p^2}{2m_B}-\mu+V_{\rm lat}(\bm r)\right] \Phi (\bm r)+\frac{g_{BB}}{ 2}
\int d^d \bm r \Phi^\dag (\bm r)\Phi^\dag (\bm r) \Phi (\bm r)\Phi (\bm r)\nonumber\\
&+\int d^d \bm r \Psi^\dag (\bm r)\left[\frac{\hat p^2}{2m_F}-\mu_{F} +V_{\rm lat}(\bm r)\right]
\Psi (\bm r)+\frac{g_{BF}}{ 2} \int d^d \bm r \Phi^\dag (\bm x)\Phi (\bm r) \Psi^\dag (\bm r)  \Psi (\bm r)
\end{align}
\end{widetext}
Here $V_{\rm lat}(\bm r)$  denotes the lattice potential, $V_{\rm lat}(\bm r)=\sum_{j=1}^3V_0\sin^2(\pi \frac{r_j}{a})$ with $a$ the lattice spacing. The coupling constants for the interactions are given by $g_{BB}=\frac{ 4\pi a_{BB}}{m_B}$, $g_{BF}=\frac{ 4\pi a_{BF}}{m_{\rm red} }$, where $m_B$ is the mass of a bosonic atom, $m_{\rm red}$ is the boson-fermion reduced mass and $a_{BB/BF}$ are the boson-boson and boson-fermion scattering lengths, respectively. $\Phi( \bm r)$ and $\Psi( \bm r)$ are the boson and the fermion field operators, respectively.

We now expand the boson field operators in the Wannier basis, $\Phi( \bm r)=\sum_{i,\alpha} w_{\alpha}(\bm r-\bm r_i) b_{i,\alpha} $, where $w_{\alpha}(\bm r-\bm r_i)$ are the eigenstates of the single particle Hamiltonian $H^B_0=\hat T_B+V_{\rm lat}(\bm r)$ with $\hat T_B=\frac{\hat p^2}{2m_B}-\mu$. 
The indices $\alpha$ (and all other Greek indices used below) in 3D denote $\alpha=({\alpha_x,\alpha_y,\alpha_z})$ with $\alpha_{x,y,z}=1,2,3$ labeling 3D vibrational modes. We consider below the dynamics of the bosons moving in the lowest few Bloch bands. For $V_0 \gg E_R$, where $E_R=\pi^2/2m_B a^2$ is the recoil energy and $V_0$ is the strength of the lattice potential, the Wannier wave functions $w_{\alpha}(x)$ can be locally approximated by the wave functions of the $\alpha$th excited state of a harmonic oscillator. Here we assume that lowest lying bands are well separated from each other which allows one to uniquely define optimally localized Wannier functions\cite{Kohn,Jakschann}, \emph{i.e} $w_{\alpha}(\bm r)$ are real symmetric or antisymmetric wave functions decaying exponentially with $\bm r$. The fermion fields $\Psi( \bm r)=\sum_{i} u_{\alpha}(\bm r -\bm r_i) c_{i,\alpha}$, where the fermion Wannier wavefunctions $u_{\alpha}(\bm r-\bm r_i)$ are chosen using the mean-field Hamiltonian for the fermions, $\hat T_F+V_F(\bm r)$ with the effective mean-field potential for the fermions $V_F(\bm r)=V_{\rm lat}(\bm r)+\frac{g_{BF}}{2}\rho_B(\bm r)$. Here, $\rho_B(\bm r)=n_0 |w_{i,1}(\bm r)|^2$ with $\rho_B(\bm r)$ and $n_0$ being the average boson density per site and average boson number per site, respectively. Thus, the shapes of these functions within a unit cell depend on the sign of the interspecies interactions. Restricting the fermions to a
single band,~\cite{single-band approximation}
we get the following multi-band Hamiltonian for Bose-Fermi mixtures,
\begin{align}\label{eq:Bose-Hubbard-two-band-app}
\!H&=H_l+H'_l+H_t+H_{BF}+H_F\\
\!H_l\!&\!=\!\sum_{i,\alpha} \varepsilon_{\alpha}\hat{n}_{Bi,\alpha}+\sum_i\frac{U_{BB}}{2}\hat{n}_{Bi,1}(\hat{n}_{Bi,1}\!-\!1)\\
\!H'_l \!&\!=\!\frac{1}{2}\sum_{i,\mu \nu \sigma \lambda}\!\!\,\!\!^{^\prime} M_{\mu \nu \sigma \lambda} b^{\dag}_{i,\mu}b^{\dag}_{i,\nu }b_{i,\sigma }b_{i,\lambda}\label{eq:Hprime}\\
\!H_t\!&\!=\!-\!\!\!\sum_{<ij>,\alpha}\!\!t^{(\alpha)}\!\left[b^{\dag}_{i,\alpha} b_{j,\alpha}
\!+\! \rm H.c.\!\right]\!\!\\
H_{BF}\!\!&=\!\sum_{i,\alpha} U^{\alpha,\alpha}_{FB}[\hat n_{i,\alpha}\!-\!\langle \hat n_{Bi,\alpha}
\rangle] \hat n^F_{i}\\
\!H_F\!&=\!\sum_{<ij>}\left[\epsilon_0\hat{n}^F_{i}\delta_{ij}-t_F\left(c^{\dag}_{i} c_{j}
\!+\!H.c.\right)\right].
\end{align}
Here the boson single-particle and hopping energies are $\varepsilon_{\alpha}=\langle w_{i,\alpha}|H^B_0|{w_{i,\alpha}}\rangle-\mu$ and $t^{(\alpha)}=-\langle{w_{i,\alpha}} |\hat{T}_B+V_{\rm lat}(\bm r)|{w_{j,\alpha}}\rangle$, respectively. The boson-boson and boson-fermion interactions are given by the following overlap integrals: $U_{BB}=g_{BB}\langle{w_{i,1};w_{i,1}}\!| {w_{i,1};w_{i,1}}\rangle$ and $U^{\alpha, \alpha}_{FB}=g_{BF}/2\langle{w_{i,\alpha};u_{i}}| {w_{i,\alpha};u_{i}}\rangle$.The fermion energy and hopping are, $\epsilon_0=\langle{u_{i}}| \hat{T}_F+V_F(\bm r)|{u_{i}}\rangle-\mu_F$  and $t_F=-\langle{u_{i}} |\hat{T}_F+V_F(\bm r)|{u_{j}}\rangle$, respectively.
The matrix elements $M_{\mu \nu \sigma \lambda}$ are defined as $M_{\mu \nu \sigma \lambda}=\frac{g_{BB}}{2}\langle{w_{i,\mu};w_{i,\nu}}\!| {w_{i,\sigma};w_{i,\lambda}}\rangle$.
In the lowest band, $U_{BB}=M_{1 1 1 1}$. The prime in the summation sign on the right-hand-side of Eq.~\eqref{eq:Hprime} indicates that $M_{1 1 1 1}$
is excluded from the sum.

Instead of treating the full, complex, multi-band Hamiltonian, \cite{Scarola_PRL05, Isacsson_PRA05, Liu_PRL06} we consider, for simplicity, a three-band model for the bosons and treat the fermions within a single band. We also keep only the largest band-mixing terms in the bosonic part of the Hamiltonian.~\cite{Lutchyn_PRA09} The effective three-band model is justified for large inter-band energy
separation $\Omega=\sqrt{4E_RV_0}$. In Ref.~[\onlinecite{Lutchyn_PRA09}], we show that the band-mixing processes coupled with the fermion contact interactions,
within an effective two band model for the bosons, lead to an increase of boson-boson repulsion.
Here, using an effective three-band model, we will show that the band-mixing processes additionally lead to a significant renormalization of the bosonic hopping parameter as well. Thus, the three-band effective model is the minimal model that captures all the significant multi-band effects. The band-mixing processes involving the higher bands ($\alpha=4,5,...$) have the same qualitative effects on the low-energy Hamiltonian. Therefore, in the rest of the paper we restrict our analysis to the lowest three bands $\alpha=1,2,3$.

\subsection{Renormalization of boson-boson interaction}

The renormalization of boson-boson interaction appears due to the presence of the band-mixing terms given by $H_l'$ in Eq.~\eqref{eq:Bose-Hubbard-two-band-app}. In addition to the scattering of bosons within the lowest band given by the energy $U_{BB}$, there are processes involving scattering of bosons to higher Bloch bands. The particle(s) promoted to higher Bloch bands have much higher energy set by $\Omega$, and, thus, can stay in these excited states for a short time $\sim 1/\Omega$. Because of a large gap $\Omega \gg U_{BB},\mu, t$, higher Bloch bands can be integrated out yielding the renormalization of the parameters of the single-band Bose-Fermi model. Within the effective three-band model, the dominant band-mixing processes correspond to scattering of two bosons from the first band to two bosons in the second band described by the amplitude $M_{2211}$ and scattering of two bosons from the first band to the first and the third bands given by $M_{3111}$.~\cite{band-mixing} In order to calculate corrections to the boson Hamiltonian due to the band-mixing terms $H_l'$, we use a perturbation series in $1/\Omega$. The lowest order corrections to $U_{BB}$ appear in the second order and are proportional to $M_{2211}^2/\Omega$ and  $M_{3111}^2/\Omega$. However, these renormalizations are present in a pure bosonic system as well and are independent of $U_{FB}$. Although these are important corrections, we neglect them here since they are not modified by the presence of the fermions, which is the focus of the present paper. We will assume that these corrections are already taken into account, yielding the renormalized Bose-Hubbard parameters $\tilde U_{BB}$ and $\tilde t$.

 Let us now concentrate on the renormalizations of the Bose-Hubbard parameters which are proportional to $U_{FB}$. The lowest non-trivial effects appear at the third-order in a perturbation expansion:
\begin{align}
\Delta H_l=\sum_{s,s'}\frac{\langle{n_0}|H'_l|{s}\rangle\langle{s}|H_{BF}|{s'}\rangle\langle{s'}|H'_l|{n_0}\rangle}{(E_0-E_s)(E_0- E_{s'})}\ket{n_0}\!\bra{n_0}.
\end{align}
Here the state $|{n_0}\rangle$ denotes the state with all the bosons in the lowest band ($\langle \hat n_{B,1} \rangle =n_0$),
and the state $|{s}\rangle$ denotes intermediate states where one or two bosons are excited to the higher vibrational states.
The explicit evaluation of the matrix elements yields the result,
\begin{align}
\Delta H_l\!&=\!\sum_{i} \hat n_{Fi} \hat n_{Bi}(\hat n_{Bi} \!-\!1) \!\sum_{\mu \neq 1} \frac{M_{11\mu \mu}^2
(U_{FB}^{\mu \mu}\!-\!U_{FB}^{11})}{(E_{\mu}\!+\!E_{\nu})^2} \!\nonumber\\
&+\sum_{i} \hat n_{Fi} \hat n_{Bi}(\hat n_{Bi} \!-\!1)^2 \! \sum_{\mu\neq 1} \frac{M_{111\mu}^2 (U_{FB}^{\mu \mu}\!-\!
U_{FB}^{11})}{E_{\mu}^2} \!\nonumber,
\end{align}
where $E_{\lambda}=\Omega (\lambda_x +\lambda_y+\lambda_z -3)$ with $\lambda_{x,y,z}$ being the quantum numbers labeling vibrational states along $x,y,z$ directions, respectively.
\begin{figure}
\centering
\includegraphics[width=0.9\linewidth]{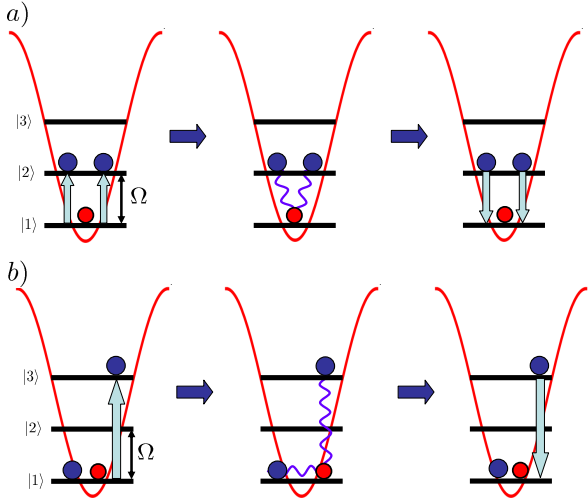}
  \caption{(color online) Virtual processes involving second (a) and third (b) Bloch bands leading to the fermion-dependent renormalization of the boson-boson interaction. Large (blue) and small (red) circles represent bosonic and fermionic atoms, respectively.} \label{fig:urenormalization}
\end{figure}
Within the effective three-band model, the correction to the boson-boson interaction becomes
\begin{align}
\Delta H_l\!&=\!\sum_{i} \hat n_{Fi} \hat n_{Bi}(\hat n_{Bi} \!-\!1)\\
&\times 3 \left[ \frac{M_{112_x 2_x}^2 (U_{FB}^{2_x 2_x}\!-\!U_{FB}^{11})}{4\Omega^2}+\frac{M_{11 3_x 3_x}^2 (U_{FB}^{3_x 3_x }\!-\!U_{FB}^{11})}{16\Omega^2}\right]\nonumber\\
&+\sum_{i} \hat n_{Fi} \hat n_{Bi}(\hat n_{Bi} \!-\!1)^2  \frac{3 M_{1113_x}^2 (U_{FB}^{3_x 3_x}\!-\!U_{FB}^{11})}{4\Omega^2}.\nonumber
\end{align}
Here the factors of $3$ come from the symmetry of the overlap integrals, i.e., $M_{112_x 2_x}=M_{112_y 2_y}=M_{112_z 2_z}$. The magnitude of the overlap integrals can be evaluated within harmonic approximation. In general, we find that these overlap integrals quickly decrease with the increase of the band number. Therefore, the largest contributions to the renormalization of the boson parameters come from the lowest excited vibrational states. Using the values of the overlap integrals, we obtain the following correction to the boson-boson interaction,
\begin{widetext}
\begin{align}\label{eq:renor_U1}
\Delta H_l\!&=\!\sum_{i} \frac{3 \tilde{U}_{BB}^2}{16\Omega^2} \hat n_{Fi} \hat n_{Bi}(\hat n_{Bi} \!-\!1)  \left[U_{FB}^{2_x 2_x}\!-\!U_{FB}^{11}+\frac{9}{64} (U_{FB}^{3_x 3_x }\!-\!U_{FB}^{11}) \right]+\frac{3 \tilde{U}_{BB}^2}{32 \Omega^2} \sum_{i} \hat n_{Fi} \hat n_{Bi}(\hat n_{Bi} \!-\!1)^2 (U_{FB}^{3_x 3_x}\!-\!U_{FB}^{11})\nonumber\\
&=-U_{BF} \frac{\tilde{U}_{BB}^2}{\Omega^2} \sum_{i} \hat n_{Fi} \hat n_{Bi}(\hat n_{Bi} \!-\!1) \left[  \frac{3}{16} p_{12}+ \left(\frac{27}{1024}+ \frac{3}{32}  \hat n_{Bi} \right)p_{13} \right].
\end{align}
\end{widetext}
Here the dimensionless parameter $p_{1\alpha}$ is given by
\begin{align}\label{eq:p}
p_{1\alpha}&=1-\frac{U_{FB}^{\alpha_x \alpha_x}}{U^{1,1}_{FB}}=1\!-\frac{\langle{w_{i,\alpha_x};u_{i}}| {w_{i,\alpha_x};u_{i}}\rangle}{\langle{w_{i,1};u_{i}}| {w_{i,1};u_{i}}}\rangle.
\end{align}
Using Eq.\eqref{eq:renor_U1}, the renormalized boson-boson interaction $U_{\rm eff}$ becomes
\begin{align}\label{eq:Ueff}
U_{\rm eff}\approx \tilde{U}_{BB}\left(1-\frac{3\tilde{U}_{BB}U_{BF}}{8\Omega^2} n_{F}\left[p_{12}+ \left(\frac{9}{64}+\frac{n_0}{2}\right)  p_{13} \right]\right).
\end{align}

To understand the effects of the fermions on $U_{\rm eff}$, we first consider the case of attractive interspecies interactions, $U_{BF} < 0$. The sign of the correction depends on the sign of $p_{1\alpha}$. Since boson and fermions attract each other, the fermion wavefunction $u(\bm r)$ becomes peaked at the center of a unit cell. Therefore, the overlaps of $u_i$ with the boson Wannier functions in the second, $w_{i,2}$, and third, $w_{i,3}$, Bloch bands are smaller than its overlap with $w_{i,1}$. Thus, for negative $U_{BF}$, $p_{12}$ and $p_{13}$ are positive and ${\cal{O}}(1)$. Consequently, the presence of fermions leads to the increase of the boson-boson repulsion.

In the opposite limit, $U_{BF} > 0$, the fermions and bosons repel each other within the unit cell.
Assuming that the number of the bosons per site is larger than that of the fermions, the fermion density is likely to be suppressed at the center of a unit cell, resulting in the numerator in the second term in Eq.~(\ref{eq:p}) exceeding the denominator, $p < 0$.
Since the sign of the renormalization to $U_{\rm eff}$ in Eq.~(\ref{eq:Ueff}) is determined by ${\rm sgn}(pU_{FB})$, the renormalization is
again repulsive, and it boosts $\tilde{U}_{BB}$ as in the case of $U_{BF} < 0$

\subsection{Renormalization of boson hopping parameter}

In addition to the renormalization of $\tilde{U}_{BB}$, the presence of the fermions leads to a renormalization of the boson hopping energy as well. The higher-band renormalizations of $t$ appear in the fourth order of the perturbation theory, and, thus, in principle, are smaller in $1/\Omega$. However, the boson tunneling rate between the nearest neighbor sites is much larger for the second and third bands compared to the first one, $t^{(1)}\ll t^{(2,3)}$. Therefore, the higher-band corrections to the hopping energy can also be significant.
\begin{figure}
\centering
\includegraphics[width=0.9\linewidth]{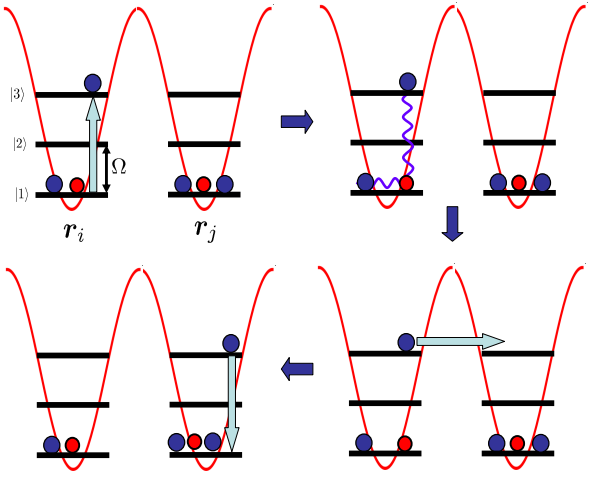}
  \caption{(Color online) Virtual processes leading to the renormalization of the bosonic hopping. Here the sequence of events corresponds to the amplitude $A^{(2)}$ (see text). Large (blue) and small (red) circles represent bosonic and fermionic atoms, respectively.} \label{fig:trenormalization}
\end{figure}

 Let us consider the tunneling process of a boson to a nearest neighbor site through the higher Bloch bands. Again, we find that there are significant corrections to hopping which are independent of the boson-fermion interactions. These virtual processes lead to the renormalization of the hopping parameter within the bare Bose-Hubbard
 model. As before, we ignore these processes below. We now consider the lowest-order, fermion-dependent, tunneling processes through the virtual states. The lowest-order corrections to the tunneling Hamiltonian are
\begin{align}\label{eq:hopping_multi}
\Delta H_t\!=\!\sum_{<ij>,m,n}\!A(n,m)\!\left\{\ket{n_i,m_j}\!\!\bra{n\!-\!1_i,m\!+\!1_j}+{\rm H.c.}\right\}
\end{align}
Here $\ket{n_i,m_j}$ is the state of the Mott insulator with $n$ and $m$ bosons on sites $i$ and $j$, respectively.
The state $\ket{n\!-\!1_i,m\!+\!1_j}= \ket{n-1}_i \otimes \ket{m+1}_j$ where the
nearest-neighbor sites $i$ and $j$ have one less and one more boson with respect to the   original state $\ket{n_i,m_j}$, and the occupation of all other sites remains the same. One can now calculate the lowest-order fermion-dependent amplitude $A(n,m)$ for the tunneling process . Assuming that the initial state and final states are $\ket{i}=\ket{n_i,m_j}$ and $\ket{f}=\ket{n\!-\!1_i,m\!+\!1_j}$, respectively, the amplitude $A(n,m)$ is given by
\begin{align}
&A(n,m)\!=\!A^{(1)}+A^{(2)},\\
&A^{(1)}\!=\!\sum_{s,s'}\frac{\bra{i}H'_{l}\ket{s}\!\bra{s}H_t \ket{s'}\!\bra{s'} H_{BF} \ket{s'}\!\bra{s'} H'_{l} \ket{f}}{ (E-E_s)(E-E_{s'})^2},\\
&A^{(2)}\!=\!\sum_{s,s'}\frac{\bra{i}H'_{l}\ket{s}\!\bra{s}H_{BF}\ket{s}\!\bra{s} H_t \ket{s'}\!\bra{s'} H'_{l} \ket{f}}{ (E-E_{s'})(E-E_{s})^2}.
\end{align}
Here $\ket{s}$ denotes the intermediate states with one or two excited bosons. The dominant contribution to the amplitude $A$ comes from the virtual processes involving the third Bloch band, see Fig~\ref{fig:trenormalization}. Virtual processes involving the second band are proportional to $(t^{(2)})^2$ since they require to transfer two bosons from site $i$ to site $j$. Therefore, their contribution to $A(n,m)$ is much smaller compared to the one from the third band, and can be neglected. The resulting expression for the amplitude $A(n,m)$ becomes
\begin{align}\label{eq:amplitude_multi}
\!A(n,m)\!&=\!-2t^{(3)}\frac{M_{3111}^2}{\Omega^3} (n-1) m \sqrt{n(m\!+\!1)} \nonumber \\&\times \left[U^{(1,1)}_{FB}\!-\!U^{(3,3)}_{FB}\right]\!n_i^F.
\end{align}
Note that the amplitude for tunneling through virtual state is zero if either $n=1$ or $m=0$ since the band-mixing processes require to have at least two bosons on the same site. In the Mott-insulating state the average number of bosons per site is $n_0$. Hence, the dominant contribution to the hopping Hamiltonian \eqref{eq:hopping_multi} comes when $m=n=n_0$. (In the Mott-insulating state, one can neglect the fluctuations of boson occupation per site at the mean field level.) Using Eqs. \eqref{eq:hopping_multi} and \eqref{eq:amplitude_multi} and evaluating the overlap integrals in the harmonic approximation, we find the renormalized boson hopping energy in the Mott-insulating state to be
\begin{align}\label{eq:ren_hopping}
\!t_{\rm eff}\!=\tilde t^{(1)}\left[\!1+\frac{t^{(3)}}{ \tilde t^{(1)}}\frac{\tilde{U}_{BB}^2(n_0\!-\!1)n_0 \sqrt{n_0(n_0\!+\!1)}}{4\Omega^3} U_{FB} p_{13} n_i^F\right]
\end{align}
with the function $p_{13}$ defined in Eq.~\eqref{eq:p}. The tunneling matrix element $t^{(3)}$ is given by,
\begin{align}
t^{(3)}&=\frac{E_R}{4}\left[M_{A} \left(3,-\frac{V_0}{4 E_R}\right)-M_{A}\left(2,-\frac{V_0}{4 E_R}\right)\right],
\end{align}
where $M_{A}(r,x)$ is the characteristic Mathieu function.\cite{stegun_book} In Eq.~\eqref{eq:ren_hopping}, the hopping energy for the lowest band $\tilde{t}^{(1)}$ includes all the corrections due to the fermion-independent virtual processes (fermion-independent higher-band correction to the hopping energy goes as $\delta t^{(1)} \propto t^{(3)} U_{BB}^2/\Omega^2$). Therefore, the self-consistent perturbation theory for the tunneling is well-defined, and the second term in the brackets in Eq.~\eqref{eq:ren_hopping} is smaller than one. However, given that $U_{BB}^2U_{FB}/\Omega^3 \ll 1$ and $t^{(3)}/ \tilde t^{(1)}\gg 1$, the renormalization of the boson tunneling is still a significant effect.

We now discuss the sign of the correction to the boson hopping parameter. As mentioned before, in the case of attractive interaction between the fermions and the bosons ($U_{FB}<0$), the function $p_{13}>0$. Therefore, the higher-band virtual transitions lead to a suppression of the boson hopping parameter. This can be also understood using the arguments in Ref.~[\onlinecite{luhmann-2007}]: the presence of fermions leads to the squeezing of the bosonic Wannier wavefunction, and, thus, should reduce the bosonic hopping energy.
In the case of repulsive inter-species interaction ($U_{FB}>0$), the bosons and the fermions like to maximize the distance between each other. Assuming, as a result, that the fermion density is substantially suppressed at the center of the unit cell, the function $p_{13}$ becomes negative. Thus, the presence of the fermions leads to the suppression of the bosonic hopping energy in this case as well.

\subsection{Shift of the phase diagram}

We now discuss the effects of the fermions on the phase boundary between
the Mott insulator and the superfluid states. Within the effective three-band model,
we find that the parameters of the bare Bose-Hubbard model are renormalized as $\tilde{U}_{BB}\rightarrow U_{\rm eff}$ and
$\tilde{t}\rightarrow t_{\rm eff}$ according to Eqs.~\eqref{eq:Ueff} and \eqref{eq:ren_hopping}. The calculation of the phase boundary can be done in a similar manner as in Sec.\ref{section:4}.
Integrating out the fermions, and neglecting terms of the order of $U^2_{FB}/\Omega^4$, leads to the effective imaginary-time action analogous to Eq.~(\ref{eq:eff-action}),
\begin{align}\label{eq:eff_action}
\!\!S_{\rm eff}(b^{*}, b)\!&\!=\!\sum_i\!\int_{0}^{\beta}\!\! d\tau \! \left[b^{*}_i \partial_{\tau} b_i\!+\! \frac{U_{\rm eff}}{2}\hat{n}_{Bi}(\hat{n}_{Bi}\!-\!1)
\!-\! \tilde \mu \hat{n}_{Bi}\!\right]\nonumber\\
\!&\!-\! \int_{0}^{\beta}\!\! d\tau\!  \sum_{<ij>}\!\!t_{\rm eff}\!\left(b^{\dag}_{i} b_{j}
\!+\! b^{\dag}_{j} b_{i}\!\right) \\
&-\!\sum_{ij}\!\! \int_0^{\beta}\! d \tau_1\! \int_0^{\beta} d \tau_2
n_{Bi}(\tau_1)M_{ij}(\tau_1\!-\!\tau_2) n_{Bj}(\tau_2)\nonumber.
\end{align}
Here $U_{\rm eff}=\tilde U_{BB} + \delta U_{BB}$, $t_{\rm eff}=\tilde t + \delta t$, $\tilde \mu=\mu -U_{FB}n^0_{Fi}$, with $n^0_{Fi}$  the average density of the fermions. The change in boson-boson interaction $\delta U_{BB}$ and bosonic hopping $\delta t$ are given by
\begin{align}
\delta U_{BB}&=-\frac{3\tilde U_{BB}^2 U_{BF}}{8\Omega^2} n^0_{F}\left[p_{12}+ \left(\frac{9}{64}+\frac{n_0}{2}\right)  p_{13} \right],\\
\delta t&=t^{(3)}\frac{\tilde U_{BB}^2(n_0\!-\!1)n_0 \sqrt{n_0(n_0\!+\!1)}}{4\Omega^3} U_{FB} p_{13} n_i^F.
\end{align}
The last term in Eq.~(\ref{eq:eff_action}) describes the screening of the bosonic repulsive interactions by the fermions, same as in the single-band model, leading to the suppression of the Mott insulating phase.

\begin{figure}
\centering
\includegraphics[width=0.99\linewidth]{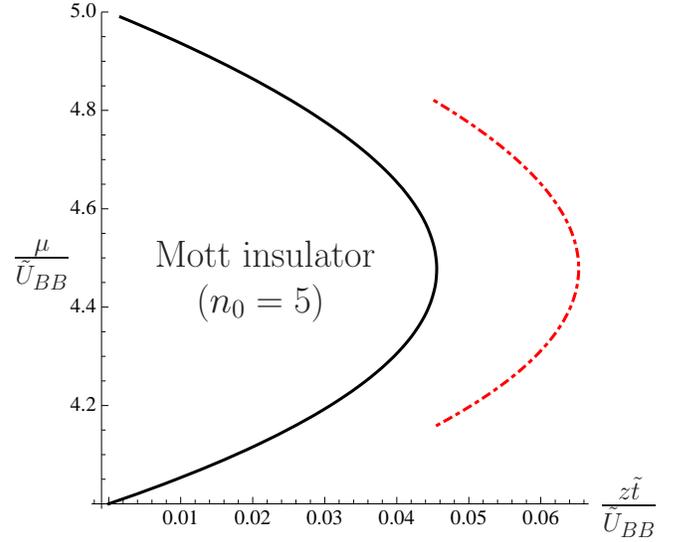}
  \caption{(Color online) Phase boundary of the boson Hubbard model in 3D with and without the fermions. Here we choose the boson density $n_0=5$ for illustration purposes. Solid line describes the insulator-superfluid phase boundary without the fermions. The dashed line corresponds to the phase boundary in the presence of the fermions. Here we used the parameters specified below Eq.\eqref{eq:estimates} and $U_{BF}=\tilde U_{BB}/2$.} \label{fig:phase_boundary_multi}
\end{figure}

The effect of the two competing contributions -- fermion screening and the effects of the higher bands -- on the phase diagram can be calculated analytically using the framework developed in the
preceding sections. We again need to calculate the boson on-site Green's function for the action in Eq.~(\ref{eq:eff_action})
at zero frequency,
\begin{align}\label{eq:Green's_mb}
\!&\!\!{{G}}^{3B}_i(\omega_n=0)\!=\\
\!&\!-\!\frac{n_0\!+\!1}{\delta E_p}\!\left[\!1\!-\!\frac{\delta U_{BB} n_0}{\delta E_p}+\!\frac{U_{FB}^2}{\Delta \delta E_p}\!R^{}\!\left(\!\frac{\delta E_p}{4E_F} \! \right) \! \right]\nonumber\\
\!&\!-\!\frac{n_0}{\delta E_h}\!\left[\!1\!+\!\frac{\delta U_{BB} (n_0-1)}{\delta E_h}+\!\frac{U_{FB}^2}{\Delta \delta E_h}\!R^{}\!\left(\!\frac{\delta E_h}{4E_F} \! \right) \! \right]\!.\nonumber
\end{align}
Here $\delta E_p$ and $\delta E_h$ are the new, renormalized, particle and hole excitation energies: $ \delta E_p={\tilde U_{BB} n_0 -\tilde \mu}$ and $\delta
E_h={\tilde \mu -\tilde U_{BB} (n_0 -1)}$, where $n_0$ is the number of bosons per site minimizing the ground state energy.

As before, the mean field
superfluid-insulator phase boundary can be obtained by solving the new equation, analogous to Eq.~(\ref{eq:r}),
\begin{align}\label{eq:r_mb}
\frac{1}{z t_{\rm eff}}+\int_{-\beta}^{\beta} d\tau {{G}}^{3B}_i(\tau)=0.
\end{align}
The shift of the superfluid-insulator phase boundary can be obtained by looking at the change of the critical hopping $t_c$ where the transition occurs with and without fermions, $\delta t_c=t_c(U_{BF})\!-\!t_c(U_{BF}\!=\!0)$. To the linear order in $U_{FB}$, the correction to the phase boundary is given by,
\begin{align}\label{eq:tc1}
\frac{\delta t_c^{(1)}}{\tilde U_{BB}}&=\!-\frac{\delta t}{\tilde{U}_{BB}}+\frac{\delta U_{BB}}{z\tilde U_{BB}} \\  &\times\left[\frac{[1+2(\tilde \mu/ \tilde U_{BB})^2-2(\tilde \mu/\tilde U_{BB})(n_0-1)-n_0] n_0}{[1+(\tilde \mu/\tilde U_{BB})]^2}\right]\nonumber
\end{align}
Notice that, as discussed before, the product of $U_{BF}$ and $p$ remains negative, \emph{irrespective of the sign of $U_{BF}$}. Therefore, $\delta t <0$ and $\delta U_{BB}>0$.
For $\tilde{\mu}/\tilde{U}_{BB}$, we use the value near the tip of the Mott lobes in the bare model (without the fermions):
$\tilde{\mu}/\tilde{U}_{BB}|_{\rm tip}=\sqrt{n_0(n_0+1)}-1$.  By substituting this value into Eq.\eqref{eq:tc1} one finds that
the shift $\delta t_c^{(1)}$ is given by
\begin{align}\label{eq:tc1tip}
\left. \frac{ \delta t_c^{(1)}}{\tilde U_{BB}} \right|_{\rm tip}=-\frac{\delta t}{\tilde U_{BB}}+\frac{\delta U_{BB}}{z \tilde U_{BB}} \left( 1 + 2 n_0 - 2 \sqrt{ n_0 (1 + n_0) }\right).
\end{align}
One can see that the above is always positive, indicating that the Mott-insulating phase expands at the tip of the lobes.

We now consider the effect of the fermion-mediated screening, which manifests only in the second order in $U_{FB}$.
 The shift of the critical hopping, $\delta t^{(2)}_c$, is given by,
\begin{align}\label{eq:Green's_dyn_mb}
& \frac{\delta t_c^{(2)}}{\tilde U_{BB}}\!=- {U_{FB}^2 \over  z \Delta \tilde U_{BB} \left(1+\frac{\tilde \mu}{\tilde U_{BB}}\right)^2}\!  \\
&\times \left \{ {\!(n_0\!+\!1)\!\left(\frac{\tilde \mu} {\tilde U_{BB}}\!-\!(n_0\!-\!1)\!\right)^2}\! R\!\left(\!\frac{\tilde U_{BB}}{4E_F}\left[n_0\!-\!\frac {\tilde \mu} {\tilde U_{BB}} \right]\right) \right. \nonumber \\
& \left. \!+ {n_0 \left(n_0\!-\! \frac{\tilde \mu}{\tilde U_{BB}}\right)^2} R\!\left(\!\frac{\tilde U_{BB}}{4E_F}\!\left[\frac {\tilde \mu} {\tilde U_{BB}} \!-\!(n_0\!-\!1)\!\right]\! \right) \! \right \}\!, \nonumber
\end{align}
which indicates that $\delta t^{(2)}_c< 0$, and, thus, the fermion-mediated screening leads to the suppression of the Mott insulating phase. At the tip of Mott lobes,  $\delta t^{(2)}_c$ becomes
\begin{align}\label{eq:Green's_dyn_mbtip}
 \frac{\delta t_c^{(2)}}{\tilde U_{BB}}\!&=- {U_{FB}^2  \left( 1 + 2 n_0 - 2 \sqrt{ n_0 (1 + n_0) }\right) \over  z \Delta \tilde U_{BB}}\!  \\
\!&\!\times\!\! \left \{\!R\!\left(\!\frac{\tilde U_{BB} [\sqrt{n_0(1\! +\! n_0)}\!-\!n_0]}{4E_F}\!\right)\!+\right.\\
& \left. + R\!\left(\!\frac{\tilde U_{BB} [1 \!+\! n_0 \!-\! \sqrt{n_0(1\! +\! n_0)}]\!}{4E_F}\!\right) \! \right \}\!. \nonumber
\end{align}

We now compare the two effects, multi-band and fermion-mediated screening, for $n_0 \gg 1$.  At the tip of the Mott lobes, the two corrections to the phase boundary come with a ratio,
\begin{align}\label{eq:estimates}
\left |\frac{\delta t_c^{(1)}}{\delta t_c^{(2)}}\right|&\sim\frac{-4 z n_0 \delta t+\delta U_{BB}}{{U_{FB}^2 \over \Delta} R(\frac{\tilde U}{8E_F})} \\
&\approx \left(4zn_0^3 \frac{t^{(3)}}{\Omega}+\frac{3}{16}\right) \frac{ \tilde U_{BB}^2 \Delta n_0 n_0^F |p_{13}|}{\Omega^2 |U_{BF}| R(\frac{\tilde U_{BB}}{8E_F})}
\end{align}
For the purposes of illustration, we choose some typical experimental parameters: $n_0=5$, $z=6$, $n^0_F=0.75$, $E_R=2.6\cdot 10^{-30}$J, $V_0/E_R=9$, $\Omega/E_R=6$, $\tilde U_{BB}/E_R=0.2$, $t^{(3)}/E_R=0.9$, $\Delta/E_R=47$ and $E_F/E_R=0.8\cdot 10^{-2}$. Assuming $U_{BF}<0$, one can estimate $p_{13}\approx 0.6$. For these parameters the function $R(\frac{\tilde U_{BB}}{8E_F})\approx 0.02$, and the ratio $| \delta t_c^{(1)}/\delta t_c^{(2)} |$ becomes $| \delta t_c^{(1)}/\delta t_c^{(2)} | \sim \tilde U_{BB} / |U_{BF}| \cdot 10^3$ indicating that for realistic parameters the higher-band effect is dominant. Thus, we conclude that the superfluid state is suppressed for either sign of the interspecies interaction, as shown in Fig.~\ref{fig:phase_boundary_multi}.

\section{Summary and conclusion}
\label{section:6}

In this paper, we have addressed the quantum phase diagram of the Bose-Hubbard
model in the presence of a degenerate gas of spin-polarized fermions with the bosons and the fermions interacting via contact interactions.
We have addressed this question by developing a framework for carrying out perturbation theory in the Bose-Hubbard model.
We first conclude that, for the single-band Bose-Hubbard model, the degenerate gas of fermions
intrinsically shrinks the area occupied by the Mott
insulating lobes (Fig.~\ref{fig:phase_boundary}).
For the multi-band Bose-Hubbard model, which is more general and experimentally relevant,  we show that the virtual transitions of the bosons to
the higher Bloch bands, coupled with the contact interactions with the fermions,
  generate a new renormalization of the interaction and the hopping parameter of the bosons. For either sign of the coupling between the fermions and
the bosons, this renormalization enhances the boson on-site
repulsion and decreases the hopping parameter, favoring the Mott insulating phase.
  If this effect is dominant over the fermion mediated screening effect, the superfluid coherence of the
  Bose-Hubbard system will be suppressed by the fermions, irrespective of the sign of the interspecies interaction, as has been
  observed in recent experiments. ~\cite{gunter_PRL06,ospelkaus_PRL06,Bloch08}
  Thus we conclude that the reconciliation of the experimental observations with the theory of
  the Bose-Hubbard model lies in the higher-band effects, which is also supported by the numerical calculations
  in Ref.~[\onlinecite{luhmann-2007}]. Note that, using our general analytical framework, we have been able to
  explain the loss of superfluid coherence for either sign of the Bose-Fermi interaction, while the calculations
  of Ref.~[\onlinecite{luhmann-2007}] apply only to the case when the interspecies interaction is attractive.

  We have not discussed in this paper the fluctuation effects on the phase diagram, \cite{Scalettar_PRL91, Krauth_PRL91, Freericks}
  and the fate of the universality class of the superfluid to insulator quantum phase transition in the presence of the fermions. \cite{Kun}
  Based on tree level arguments, Ref.~[\onlinecite{Kun}] argues that the critical properties of the transition at the
  tip of the Mott insulating lobes, where the dynamic critical exponent $z=1$ (the coupling constant $c_1=0$ in Eq.~(\ref{eq:Phi4})),
  are strongly modified by the fermions. On the other hand, for a generic point on the boundary of the lobes, where $z=2$, the critical
  fixed points are stable with respect to the fermion mediated interactions. The actual fate of the $z=1$ fixed point, however, is an open question,
  and to answer this one has to go beyond the tree level of the renormalization group analysis.
  It is also important to note that the fermions mediate a time- and space-dependent density-density interaction among the bosons.
  Such spatially non-local interactions may give rise to spatially ordered states, which may or may not coexist with superfluidity.
  If they do coexist with superfluidity, there may an interesting possibility of inducing bosonic supersolids in some parameter
  regime by tuning the boson-fermion interaction.

  In conclusion, we develop a perturbation theory for the Bose-Hubbard model, which is different from
  the standard theory for bosonic systems, \cite{Fetter_book} but is appropriate for the Bose-Hubbard model with
  a quartic (Hubbard-$U$) term. Using this theory, and by including the effects of the higher boson Bloch bands in a multi-band formulation, we
  explain the observed, unexpected, expansion of the Mott insulating lobes of the Bose-Hubbard model by introducing fermions.
  Our three new important findings are: (1) There are independent physical contributions to the bosonic quantum phase diagram in the presence of fermions, namely, the modification of bosonic interaction due to screening and the interband virtual transition processes in the multiband case; (2) the multiband processes are independent of the sign of the interaction, and if dominant, would lead to suppression of superfluidity, as observed experimentally, independent of the sign of the fermion-boson interaction;
  (3) the renormalization of the bosonic hopping term by the multiband transitions appears to be the most important process quantitatively.

\section{Acknowledgements}
We thank K.~Yang, G.~Refael, E.~Demler, T.~Porto, T.~Stanescu, E.~Hwang, L.~Cywinski, W.~Phillips, S.~Sachdev, C. W. Zhang, and V. Scarola for stimulating discussions. RL acknowledges
the hospitality of the National University of Kyiv-Mohyla Academy. This work is supported by ARO-DARPA, DOE/EPSCoR Grant \# DE-FG02-04ER-46139, and  Clemson University start up funds.

\appendix

\section{Calculation of the correlation function $K_{ijl}(\tau, \tau_1, \tau_2)$}
\label{app:correlation}
In this Appendix we illustrate how we calculate the correlation function $K_{ijl}(\tau, \tau_1, \tau_2)$, and show that the terms that can give divergent contributions to Eq.~(\ref{eq:Green's}) exactly cancel out. To illustrate the cancelation, we consider the correlation function $K_{ijl}(\tau, \tau_1, \tau_2)$ defined in Eq.~(\ref{eq:K-def}). Given that the on-site part of the boson Hubbard Hamiltonian conserves the number of bosons, the correlation function $K_{ijl}(\tau, \tau_1, \tau_2)$ can be calculated in the second quantized representation by inserting the representation of unity as a sum over particle-number eigenstates. Doing this, we find that the second term in Eq.(~\ref{eq:K-def}) at zero temperatures is given by
\begin{align}\label{eq:K_second}
&\langle T_{\tau}
b_i(\tau)b^{\dag}_i(0) \rangle \langle  T_{\tau} n_{j}(\tau_1) n_{l}(\tau_2) \rangle=  \\
&=\!e^{-\delta E_p \tau}n_0^2(n_0\!+\!1)\Theta(\tau)\Theta(\delta E_p)+\!e^{\delta E_h \tau}n_0^3\Theta(-\tau)\Theta(\delta E_h)\nonumber,
\end{align}
where $n_0$, $\delta E_p$ and $\delta E_h$ were defined in the text. To calculate the first term in Eq.~(\ref{eq:K-def}), we note it is important to distinguish between the cases where the indices $j$ and $l$ are the same or different from the index $i$, because they give rise to different matrix elements. To carefully take this into account, we separate the sums over $i,j$ as $\sum_{i,j}=\sum_{ij}[(1-\delta_{ij})(1-\delta_{il})+(1-\delta_{ij})\delta_{il}+(1-\delta_{il})\delta_{ij}+\delta_{ij}\delta_{il}]$. After taking care of the imaginary time orderings between the time indices $0, \tau_1, \tau_2, \tau$ and calculating the matrix elements, we find that the first term of Eq.~(\ref{eq:K-def}) is given by,
\begin{align}\label{eq:K_first}
\!&\!\langle T_{\tau}b_i(\tau)b^{\dag}_i(0) n_{j}(\tau_1) n_{l}(\tau_2) \rangle=\\
\!&\!=\! e^{-\delta E_p \tau}\Theta(\tau)\Theta(\delta E_p)\!\!\sum_{i=1..12}\! I_i\!+\!e^{\delta E_h \tau}\Theta(-\tau)\Theta(\delta E_h)\!\!\sum_{i=1..12}\!\!J_i.\nonumber
\end{align}
Here, the functions $I_i$ are given by,
\begin{align}\label{eq:I_i}
I_1&=\Theta (\tau_2\!-\!\tau_1)\Theta (\tau_1\!-\!\tau)\Theta (\tau_2)\Theta (\tau_1)n_0^2(n_0\!+\!1)\\
I_2&=\Theta (\tau_1\!-\!\tau_2)\Theta (\tau_2\!-\!\tau)\Theta (\tau_2)\Theta (\tau_1)n_0^2(n_0\!+\!1)\nonumber
\end{align}
\begin{align}
I_3&\!=\! \Theta (\tau_2\!-\!\tau)\Theta (\tau\!-\!\tau_1)\Theta (\tau_2)\Theta (\tau_1)\times \nonumber\\
&\times\left[n_0^2(n_0\!+\!1)\!+\!\delta_{ij}n_0(n_0\!+\!1)\right]\nonumber\\
I_4&=\Theta (\tau_1\!-\!\tau)\Theta (\tau\!-\!\tau_2)\Theta (\tau_2)\Theta (\tau_1)\times \nonumber\\
&\times\left[n_0^2(n_0\!+\!1)\!+\!\delta_{il}n_0(n_0\!+\!1)\right]\nonumber\\
I_5&=\Theta (\tau\!-\!\tau_1)\Theta (\tau\!-\!\tau_2)\Theta(\tau_1-\tau_2)\Theta (\tau_2)\Theta (\tau_1)\times \nonumber\\
&\times \left[n_0^2(n_0\!+\!1)\!+\!(n_0\!+\!1)n_0(\delta_{ij}\!+\!\delta_{il})\!+\!(n_0\!+\!1)\delta_{ij}\delta_{il}\right]\nonumber\\
I_6&=\Theta (\tau\!-\!\tau_1)\Theta (\tau\!-\!\tau_2)\Theta(\tau_2-\tau_1)\Theta (\tau_2)\Theta (\tau_1)\times \nonumber\\
&\times \left[n_0^2(n_0\!+\!1)\!+\!(n_0\!+\!1)n_0(\delta_{ij}\!+\!\delta_{il})\!+\!(n_0\!+\!1)\delta_{ij}\delta_{il}\right]\nonumber\\
I_7&=\Theta (\tau_2\!-\!\tau)\Theta (\tau\!-\!\tau_1)\Theta (\tau_2)\Theta (\!-\!\tau_1)n_0^2(n_0\!+\!1)\nonumber\\
I_8&=\Theta (\tau_1\!-\!\tau)\Theta (\tau\!-\!\tau_2)\Theta (\!-\!\tau_2)\Theta (\tau_1)n_0^2(n_0\!+\!1)\nonumber\\
I_9&=\Theta (\tau\!-\!\tau_1)\Theta (\!-\!\tau_2)\Theta (\tau_1)\left[n_0^2(n_0\!+\!1)\!+\!\delta_{ij}n_0(n_0\!+\!1)\right]\nonumber\\
I_{10}&=\Theta (\tau\!-\!\tau_2)\Theta (\tau_2)\Theta (\!-\!\tau_1)\left[n_0^2(n_0\!+\!1)\!+\!\delta_{il}(n_0\!+\!1)n_0\right]\nonumber\\
I_{11}&=\Theta (\!-\!\tau_2)\Theta (\tau_1-\!\tau_2)\Theta (\tau)n_0^2(n_0\!+\!1)\nonumber\\
I_{12}&=\Theta (\!-\!\tau_2)\Theta (\tau_2-\!\tau_1)\Theta (\tau)n_0^2(n_0\!+\!1)\nonumber,
\end{align}
and the functions $J_i$, which are of similar form but calculated for the other 12 time orderings corresponding to $\theta(-\tau)$, are
omitted here for simplicity.

  It is straightforward to check that the terms in Eq.~(\ref{eq:I_i}) which are independent of the site indices $i,j$ can be summed to yield $n_0^2(n_0+1)$. Thus, by comparing Eqs.~(\ref{eq:K_second}) and (\ref{eq:K_first}) one can see that these terms, which would have given divergent contributions to the Green's function (see Eq.~(\ref{eq:Green's})) cancel out leading to the $\tau>0$ part of Eq.~(\ref{eq:K-calc}.  In a similar way, for $\tau<0$, the terms in $J_i$ which are independent of the site indices sum up to $n_0^3$ and exactly cancel with the corresponding term in Eq.~(\ref{eq:K_first}), leading to the $\tau<0$ part of Eq.~(\ref{eq:K-calc}). Furthermore, since in the static approximation the effective interaction kernel $M_{ij}(\tau_1-\tau_2)$ itself imposes $\tau_1=\tau_2$, only the time orders $\Theta(\tau)\Theta(\tau_1)\Theta(\tau_2)\Theta(\tau-\tau_1)\Theta(\tau-\tau_2)$ and $\Theta(-\tau)\Theta(-\tau_1)\Theta(-\tau_2)\Theta(\tau_1-\tau)\Theta(\tau_2-\tau)$ from Eq.~(\ref{eq:K-calc}) contribute to the correlation function and one ends up with Eq.~(\ref{eq:K-calc_static}). However, for these time orders, the $\tau_1, \tau_2$ integrals are constrained to be in the interval between $0$ and $\tau$ and do not diverge. In a similar way, all the integrals in the calculation of the Green's function for the full interaction kernel are restricted to finite intervals as well, leading to finite non-zero contributions.
For the calculation of the boson Green's function with the full interaction kernel $M_{\bf{q}}(\Omega_n)$ given in Eq.~(\ref{eq:polarization3D}), we can use the fact the it is zero for either $q$ or $\Omega_n$ equal to zero. As a result the terms of Eq.~(\ref{eq:K-calc}) that contribute to Eq.~(\ref{eq:Green's}) have $i=j=l$ and the same time orders as given above. This way all the integrals in the calculation of the Green's function are restricted to finite intervals, canceling the potential divergences.



\begin{thebibliography}{17}
\bibitem{single-band approximation} This approximation is justified when the fermion mass is much smaller than that of bosons since then $\Omega \ll \Omega_F$ with $\Omega_F$ being the fermion band separation.
\bibitem{band-mixing} In addition to the band-mixing processes due to the boson-boson interactions, there are also band-mixing processes due to the boson-fermion interactions. However, the latter ones give much smaller contribution to the renormalization of the bare parameters in the limit $m_F/m_B \ll 1$ considered here.
\end{thebibliography}
\end{document}